\documentclass[a4paper,11pt]{article}

\usepackage{jheppub} 

\usepackage{multirow}
\usepackage[T1]{fontenc} 
\usepackage[dvipsnames]{xcolor}
\usepackage{placeins}
\usepackage[
backend=biber,
firstinits=true,
sorting=none
]{biblatex}
\usepackage{lineno}
\usepackage{siunitx}
\usepackage{subcaption}
\usepackage{xspace}
\title{\boldmath Track-Based Triggers for Exotic Signatures}

\author[a]{K. F. Di Petrillo}
\author[b]{J. N. Farr}
\author[c]{C. Guo}
\author[b]{T. R. Holmes}
\author[d]{J. Nelson}
\author[e]{K. Pachal}

\affiliation[a]{Fermi National Accelerator Laboratory,\\ Batavia, IL, USA}
\affiliation[b]{University of Tennessee,\\ Knoxville, TN, USA}
\affiliation[c]{Illinois Institute of Technology,\\ Chicago, IL, USA}
\affiliation[d]{Brown University,\\ Providence, RI, USA}
\affiliation[e]{TRIUMF,\\ Vancouver, BC, Canada}

\emailAdd{kdipetri@fnal.gov}
\emailAdd{tholmes@utk.edu}
\emailAdd{kpachal@triumf.ca}

\abstract{
Several compelling beyond the Standard Model scenarios predict signals that result in unconventional charged particle trajectories. Signatures for which unusual tracks are the most conspicuous feature pose significant challenges for experiments at the Large Hadron Collider (LHC), particularly for the trigger. This article presents a study of track-based triggers for a representative set of long-lived and unconventional signatures at the upcoming High Luminosity LHC. Scenarios studied include large multiplicities of low momentum tracks produced in a soft-unclustered-energy-pattern model, displaced leptons and anomalous prompt tracks predicted in a Supersymmetry model with long-lived staus, and displaced hadrons predicted in a Higgs portal scenario with long-lived scalars. Trigger efficiency is measured as a function of the baseline parameters of a track trigger, including transverse momentum and impact parameter. Recommendations for future hardware-based track triggers are presented. }

\begin{document} 
\maketitle
\flushbottom

\nolinenumbers
\setlength\parindent{0pt}
\setlength{\parskip}{0.5em}

\newcommand{\pt}{\ensuremath{p_{\mathrm{T}}}\xspace}
\newcommand{\dz}{\ensuremath{d_{0}}\xspace}
\newcommand{\absdz}{|\ensuremath{d_{0}}|\xspace}
\newcommand{\nTrack}{\ensuremath{n_{\mathrm{Track}}}\xspace}
\newcommand{\Ht}{\ensuremath{H_{\mathrm{T}}}\xspace}
\newcommand{\Lxy}{\ensuremath{L_{\mathrm{xy}}}\xspace}

\section{Introduction}

Decades of experimental tests have demonstrated the predictive power of the the Standard Model (SM). However, there are a myriad of phenomena it does not explain, including dark matter, baryon asymmetry, naturalness, and more. Many beyond the Standard Model (BSM) scenarios that address these inefficiencies predict unconventional track signatures that could be observed at high energy colliders such as the Large Hadron Collider (LHC). For example, long-lived BSM particles can travel a measurable distance before decaying, resulting in displaced track or anomalous prompt track signatures~\cite{Lee_2019}. Alternatively, models with a strongly coupled hidden valley can result in large multiplicities of soft charged particles~\cite{Knapen_2017}. These unconventional track signatures pose tremendous challenges for general purpose experiments at the LHC, particularly for the trigger.  

Due to bandwidth and storage constraints, the ATLAS and CMS experiments both rely on real-time reconstruction to select interesting collisions before detector data can be read out and stored. The system that performs this task is called the trigger. A particular challenge for the trigger is differentiating signal processes from a large background of low energy proton-proton interactions, including the multiple interactions that occur per event, referred to as pile-up. The trigger is primarily designed to select events with high momentum jets, photons, or leptons that  originate directly from a $pp$-collision of interest, or events characterized by large missing transverse momentum. These prompt signatures are easily identified in the first, hardware-based, stage of the trigger, which makes use of calorimeter and muon system information. Limited tracking information is only available at later stages of the trigger decision. 

Unfortunately, standard triggers often result in low trigger efficiency for unconventional BSM signatures. If a BSM particle decays at a distance and produces jets or leptons, the decay products will not point back to the $pp$-collision. Decay products can be mis-identified as pile-up or fail standard quality criteria. Charged BSM particles which travel a long distance before decaying can result in anomalous signals in the detector that can be confused with noise. Models with many low-energy charged particles can be difficult to differentiate from pile-up.

When searching for these unconventional BSM signatures, the most distinctive features of signal events are in the tracker. Dedicated track-based triggers are necessary to access these challenging scenarios.

\subsection{Current state of track triggers}

ATLAS and CMS currently use similar two-step trigger systems to select which events are saved for analysis. The first stage, Level 1 (L1) reduces the event rate from $40$~MHz to $100$~kHz within a latency of $\sim\SI{3}{\micro\s}$, using simple decisions implemented in custom hardware~\cite{Sirunyan:2721198}. The second stage, the High Level Trigger (HLT), uses standard software routines to reduce the event rate from $100$ kHz to $\sim 1$~kHz, with $\sim200$~ms on average to make a decision~\cite{Khachatryan:2212926,ATLAS:2020esi}.

Due to bandwidth and latency constraints, tracking information is only produced at the HLT, where it is the most CPU-intensive reconstruction process. There, limited track reconstruction routines are run in software, balancing the amount of information necessary to make a decision with the amount of time needed to reconstruct the event. For example in lepton triggers, track reconstruction is often only performed in small regions of interest that were identified by the calorimeter or muon system at L1. Full detector tracking is possible with fast algorithms and a reduced event rate. These fast algorithms result in reduced efficiency across wide ranges of track transverse momentum, \pt, and transverse impact parameter, \dz.

The upcoming High Luminosity LHC (HL-LHC) upgrades open up the opportunity to develop new track-based triggers for unconventional signatures. ATLAS and CMS both plan to install new tracking detectors and completely overhaul their trigger designs. In CMS, tracker \pt-modules will enable sufficient bandwidth reduction such that tracks with $\pt > 2$ GeV can be reconstructed at L1 \cite{CERN-LHCC-2017-009}. While the ATLAS tracker will not participate in the L1 decision at the HL-LHC, it will participate at the HLT with an increased input rate of  1 MHz. Under the baseline plan, track reconstruction will be performed at the HLT with a CPU based processing farm potentially complemented by commodity accelerator hardware \cite{Collaboration:2285584,CERN-LHCC-2020-004}. 

With additional effort, it may be possible to extend these hardware track triggers to include a range of long-lived or unconventional signatures \cite{Gershtein_2017, ftk_results}. Tracking performance is limited by detector readout rates, the hardware-based tracking technology used, output rate, and latency constraints. In general, increasing the maximum curvature, displacement, or target efficiency of reconstructed tracks increases complexity, rate, latency, or a combination thereof. In order to maintain a constant latency but extend its \dz range, a track trigger could increase its \pt threshold or decrease its target efficiency. Depending on the track trigger design, it may be possible to alter these requirements independently for prompt and displaced tracks, leaving a highly efficient prompt track trigger with a low \pt threshold, while also exploring a large range of displacements.

\subsection{Expanding track triggering for future experiments}

There has been a recent proliferation of studies on the efficacy of track-based triggers for long-lived particles and other exotic signatures, including rare Higgs decays \cite{Gershtein_2017}, soft-unclustered-energy patterns (SUEPs) \cite{Knapen_2017}, and generic displaced signatures \cite{M_rtensson_2019, https://doi.org/10.48550/arxiv.2003.03943}. ATLAS demonstrated that a hardware-based track finder algorithm could extend its range to include moderately displaced tracks \cite{ftk_results}. Displaced hardware-based tracking has been extensively studied for the HL-LHC on CMS.

However, because the signatures of these models are so varied, it is challenging to translate individual studies into an optimal design for a future trigger. In particular, for a realistic rate- and latency-limited trigger, it is unclear which efficiency trade-offs are most beneficial: is it better to expand \dz coverage, reduce \pt thresholds, or maximize overall tracking efficiency?

The answer to this question depends on the design of the trigger being considered, but it also depends on what kind of exotic signatures are being targeted. For hadronic decays, low \pt thresholds are important, but high efficiency may not be needed. For leptonic decays the reverse is true. 

The goal of this study is to provide a parameterization of trigger efficiencies for a variety of representative models, to be used as a guide for future trigger design. Specific tracking algorithms will not be discussed, but instead overall efficiency will be provided for a range of models as a function of several basic tracker coverage parameters.  

\section{Benchmark model performance}

\begin{figure}[tbp]
\centering 
\raisebox{0.39\height}{\includegraphics[width=.32\textwidth]{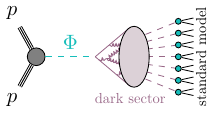}\label{subfig:suep}}
\includegraphics[width=.32\textwidth]{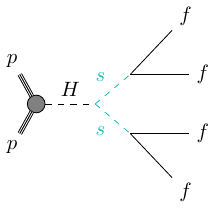}\label{subfig:higgsy}
\raisebox{0.05\height}{\includegraphics[width=.32\textwidth]{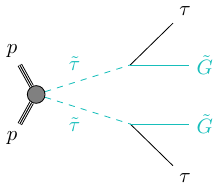}\label{subfig:stau}}
\caption{\label{fig:feynman} Feynman diagrams for benchmark processes. From left to right: SUEPs, Higgs portal, and long-lived staus. }
\end{figure}

This work studies a representative set of unconventional signatures motivated by three scenarios beyond the Standard Model . Figure~\ref{fig:feynman} shows Feynman diagrams corresponding to the benchmark models considered. Events with a large multiplicity of prompt low momentum tracks are motivated by a soft-unclustered-energy-pattern (SUEP) model. Displaced decays to hadrons are motivated by a model with long-lived scalars coupling to the Higgs boson. A gauge-mediated supersymmetry breaking scenario with direct production of long-lived staus results in displaced leptons as well as heavy (meta-)stable charged particles (HSCPs) which can be detected directly.  

Together, these four signatures span a wide range of possible track momenta and displacement, as well as event-level track multiplicity. In order to directly compare these vastly different signatures, a common strategy is used to evaluate the acceptance and efficiency of possible track-trigger configurations. The following sections outline this overall analysis strategy as well as the specific studies performed for each model.

\subsection{Analysis strategy}
\label{analysis-strat}

All studies are performed assuming HL-LHC conditions, using truth-level simulation of $pp$-collisions at a center of mass energy of $\sqrt{s}=14$~TeV. Charged particles considered for analysis are stable (or meta-stable) decay products of short-lived BSM particles, except for the HSCP case in which direct detection of the meta-stable particle is considered. For simplicity, there is no simulation of detector or material interactions, and pile-up collisions are not included in these samples.

The acceptance and efficiency of possible trigger configurations are evaluated for each signature. First, baseline tracking criteria are defined based on potential detector specifications (e.g. $\eta$ coverage, tracker radius). Only stable charged particles satisfying these criteria are considered to be in acceptance. This geometric acceptance simply asks whether particles could be reconstructed by an offline tracking algorithm with the specified detector layout, and is independent of what could then be defined by a trigger selection. Certain effects of adjusting the detector layout are explored for the HSCP signature. For other models the detector layout is kept fixed. The acceptance for a signal point is then the fraction of events with at least $n$ particles in the final state that meet the tracking criteria, where $n$ is also model dependent. 

For displaced signatures studying a \dz threshold, only tracks with $\absdz > 1$~mm are counted in the acceptance criteria. This displacement requirement separates tracks which would likely be picked up by a prompt track reconstruction algorithm from those which require a displaced tracking algorithm, and also defines a region in which SM backgrounds are small. 

Efficiency is then defined to capture the effects of track reconstruction criteria which could be set at the track-trigger level rather than enforced by the physical detector. The two most important criteria studied are the minimum \pt and maximum \absdz of tracks which would be reconstructed by the track-trigger. These factors have different effects in different BSM models, depending on whether the final state tracks are prompt or displaced, what the average decay angles of long-lived particles may be, and how much momentum the final state particles carry. To study the interplay of these effects, for a fixed acceptance, the \pt and \dz thresholds are varied and the fraction of surviving tracks measured. An event level efficiency is then defined, as for the acceptance, based on whether at least $n$ particles in the final state pass these thresholds.

In realistic tracking detectors, although the reconstruction efficiency for prompt isolated tracks is very close to $100\%$ independent of track \pt, the efficiency tends to decrease with increasing \absdz\cite{atlas_lrt}. To emulate this effect in the following studies, the likelihood of reconstructing a given track is set not as a binary by the \dz threshold (e.g. all tracks up to $\absdz = 5$~mm are reconstructed; all above 5 mm are not reconstructed), but instead as a linearly decreasing function. The likelihood of reconstructing the track begins at unity for $\dz = 0$ and decreases linearly with increasing \absdz until it reaches zero at the specified \absdz threshold.

Trigger selections in the following sections are designed such that negligible background contributions are expected in most cases. The one exception is the SUEP trigger, where background rates have not been studied but could be non-negligible due to the high pile-up expected at the HL-LHC. Background rates are also dependent on the track-trigger's ability to differentiate between pile-up vertices.

\subsection{SUEPs}

Motivated in certain hidden valley models, SUEPs represent a worst case scenario for triggers at hadron colliders~\cite{Knapen_2017,Strassler:2008bv}. In SUEP events, low momentum final state particles are not collimated into Standard Model-like jets. The most conspicuous feature is a large multiplicity of diffuse low \pt tracks. This scenario is very difficult for the trigger to distinguish from pile-up, unless the many soft tracks can be reconstructed and associated to a $pp$-collision of interest. As a result, there are currently no constraints on SUEP scenarios from the LHC.

In SUEP models, a dark SU(N) sector is connected to the Standard Model by a mediator with a mass much greater than the mass splittings amongst the dark sector particles. Energy from the production of the mediator is distributed amongst a large multiplicity of light dark hadrons, which decay back to SM particles. The 't Hooft coupling of the dark sector, $\lambda_{D} \equiv \alpha_{D} N_{C_{D}}$, where $\alpha_{D}$ is the dark coupling constant and $N_{C_{D}}$ is the number of colors, determines the angular emission of partons during showering. In SUEP scenarios this coupling is assumed to be large, and dark mesons are isotropically distributed in the mediator's rest frame.

SUEP samples are generated in \textsc{Pythia8}~\cite{Sjostrand:2014zea} using the custom plugin described in Ref.~\cite{Knapen_2017}. Events are generated assuming a scalar mediator, $\Phi$, with a range of masses: 125 GeV (matching the SM Higgs boson) and then 200 GeV to 1000 GeV in 200 GeV intervals. The SUEP shower consists of a large multiplicity of a single flavor of dark mesons with mass $m_{\pi_d} =1$~GeV and temperature $T=1$~GeV. 

In principle, dark mesons may decay to a variety of final state SM particles. Possible decay modes include pairs of leptons, hadrons, photons, and light or heavy flavor quarks. Decay products may be also be displaced, or the final state may also include undetected dark matter particles. For simplicity, this study focuses on a scenaro in which each dark meson decays promptly to a pair of light quarks, $d\bar{d}$.

\begin{figure}[htb]
     \centering
     \begin{subfigure}[b]{0.49\textwidth}
         \centering
        \includegraphics[width=\textwidth]{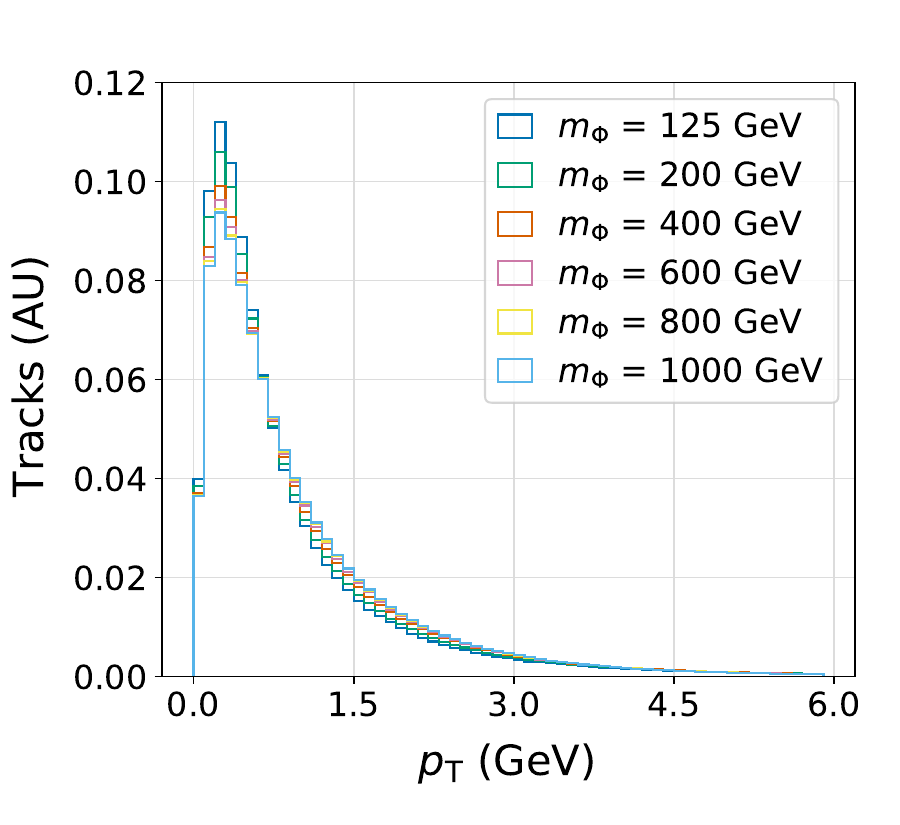}
         \caption{Charged particle \pt}
         \label{subfig:suep_pt}
     \end{subfigure}
     \hfill
     \begin{subfigure}[b]{0.49\textwidth}
         \centering
        \includegraphics[width=\textwidth]{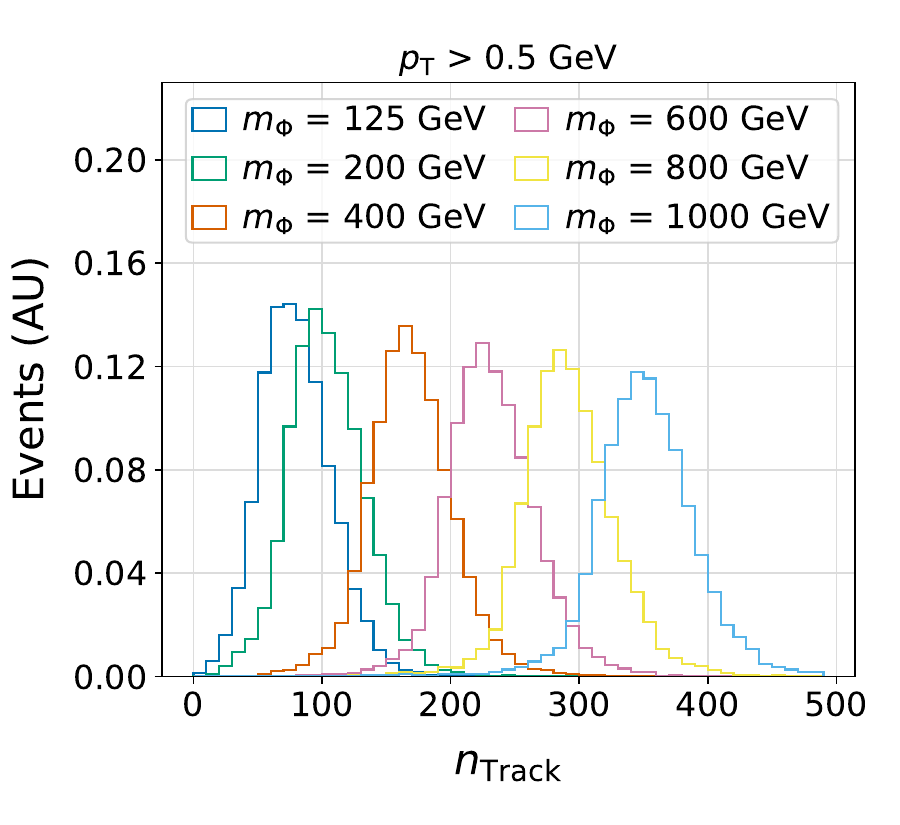}
         \caption{Charged particle multiplicity}
         \label{subfig:suep_n}
     \end{subfigure}
    \caption{Distribution of all charged SM particle transverse momenta (\subref{subfig:suep_pt}) and multiplicity of stable charged particles with $\pt > 0.5$~GeV.  (\subref{subfig:suep_n}) in SUEP events for a range of mediator masses. }
    \label{fig:suep_pt_n}
\end{figure}

Figure~\ref{subfig:suep_pt} shows charged particle transverse momenta for the SUEP benchmark models under consideration. Unlike most BSM scenarios, the \pt spectra of final state particles do not depend on the mediator mass, but on the dark meson mass and temperature. As a result, all generated momenta distributions are similar in shape, and peak sharply in the lowest \pt bins.

The number of final state particles depends on mass of the scalar mediator and dark meson mass, $n \sim m_S/m_{\pi_d}$, as shown in Figure~\ref{subfig:suep_n}. For a fixed mediator and dark meson mass, increasing the temperature reduces the number of final state particles and increases their mean momenta, as described in Ref.~\cite{Cesarotti:2020hwb}. 

Because all tracks from the model considered here are prompt, the selection strategy is relatively straightforward. The number of stable charged particles within tracker acceptance and above a given \pt threshold are counted per event. It is assumed that the track-trigger would have sufficient longitudinal impact parameter resolution to associate tracks to the correct primary vertex, and neutral final state particles are neglected.

Table~\ref{tab:suep_acceff} summarizes SUEP object-level acceptance and efficiency definitions. Tracks within $|\eta| < 2.5$ are considered to be within the detector acceptance. The event level acceptance is $100\%$ in all cases because of the large multiplicity of charged particles. The efficiency is defined as the fraction of events where the number of tracks, \nTrack, that pass the specified \pt and $\eta$ requirements is above a certain threshold.

\begin{table}[htbp]
\centering
\begin{tabular}{|c|c|}
\hline
    Variable & Requirement  \\
    \hline
    \multicolumn{2}{|c|}{Acceptance}  \\  
    \hline 
    $|\eta|$ & <  $2.5$ \\
    \hline
    \multicolumn{2}{|c|}{Efficiency}  \\
    \hline
    \pt &  $> 0.5, 1, 2$~GeV  \\
    \nTrack &  $> 100, 150, 200$  \\ 
\hline
\end{tabular}
\caption{ \label{tab:suep_acceff} Acceptance and efficiency requirements for a trigger selecting events with a high multiplicity of soft tracks}
\end{table}

\begin{figure}[htb]
     \centering
     \begin{subfigure}[b]{0.49\textwidth}
         \centering
        \includegraphics[width=\textwidth]{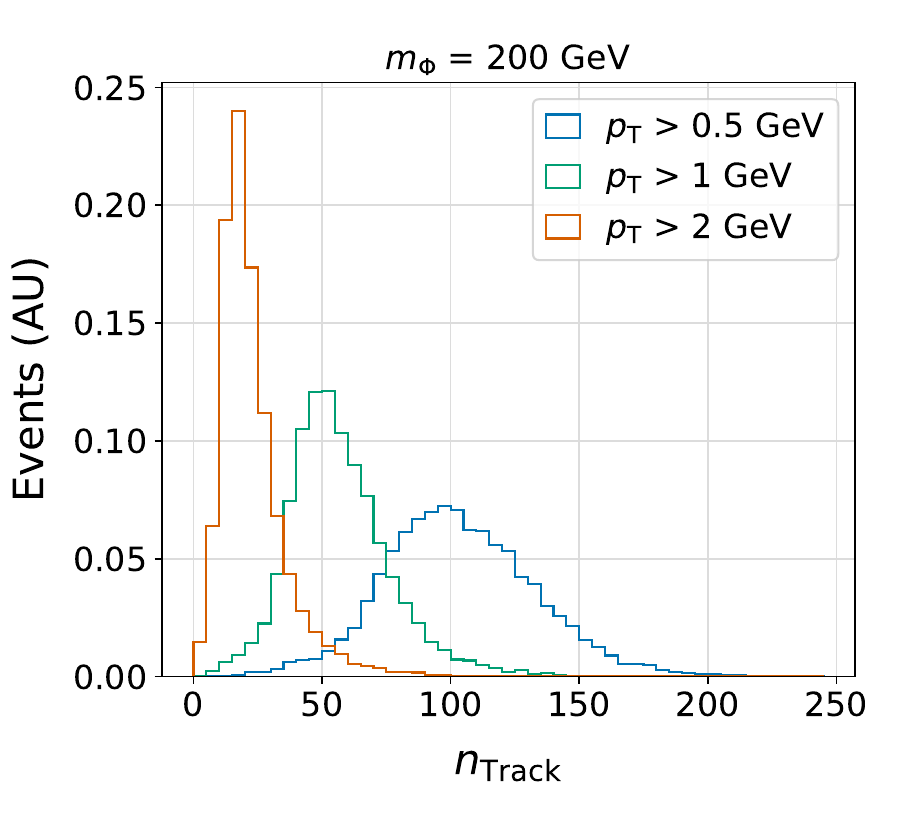}
         \caption{ }
         \label{subfig:suep_n_200}
     \end{subfigure}
     \hfill
     \begin{subfigure}[b]{0.49\textwidth}
         \centering
        \includegraphics[width=\textwidth]{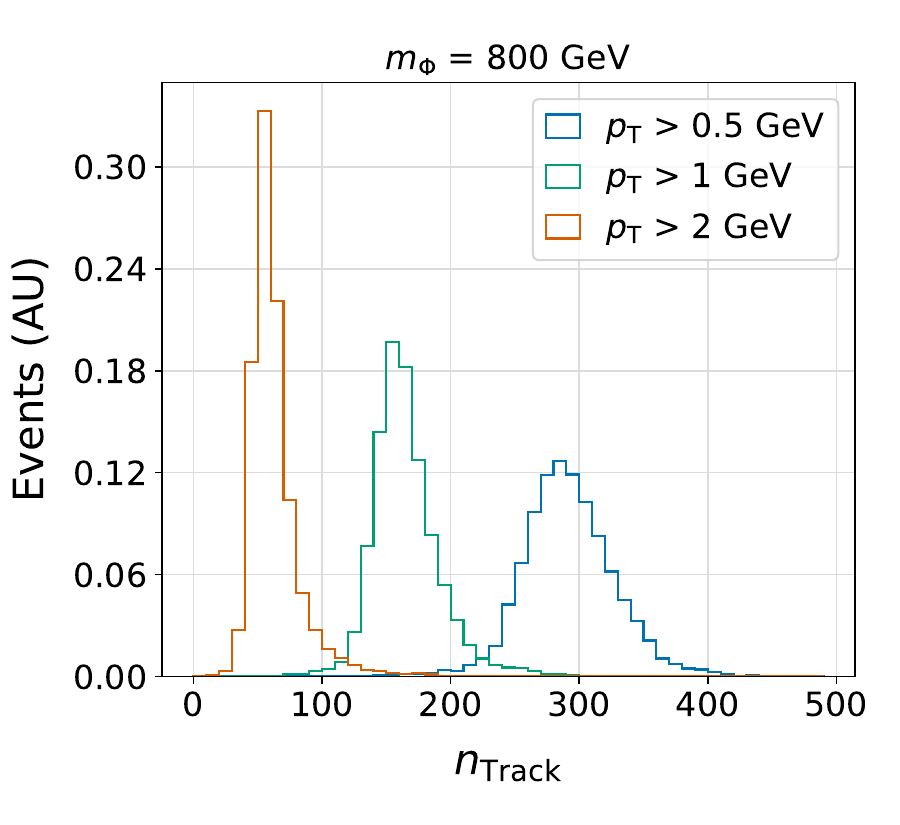}
         \caption{ }
         \label{subfig:suep_n_800}
     \end{subfigure}
    \caption{Number of charged particles per event which pass the specified track \pt threshold, for mediators of mass 200 GeV (\subref{subfig:suep_n_200}) and 800 GeV (\subref{subfig:suep_n_800}). }
    \label{fig:suep_ptcuts}
\end{figure}

The most important parameter to consider is the \pt threshold at which track reconstruction begins. For SUEP signatures, signal to background discrimination decreases with increasing track \pt requirements, because track multiplicity is the most distinctive feature of these events. This study investigates a range of \pt thresholds, from $0.5 < \pt < 2$~GeV. The lowest threshold represents the minimum charged particle \pt for most standard offline track reconstruction algorithms, aimed at charged particles that traverse a sufficient number of tracking detector layers~\cite{ATLAS_tracking_today, CMS_tracking_today}. The highest \pt threshold examined, $\pt > 2$~GeV, is based on the minimum L1 tracking threshold of 2 to 3 GeV estimated by CMS for their Phase 2 trigger upgrade~\cite{CERN-LHCC-2020-004}. For the temperature considered, a threshold above \pt $=2$~GeV, counting the \nTrack is no longer expected to be a powerful discriminator. Figure~\ref{fig:suep_ptcuts} shows the number of reconstructed tracks in SUEP events above different track \pt thresholds. 

\begin{figure}[htb]
     \centering
     \begin{subfigure}[b]{0.49\textwidth}
         \centering
        \includegraphics[width=\textwidth]{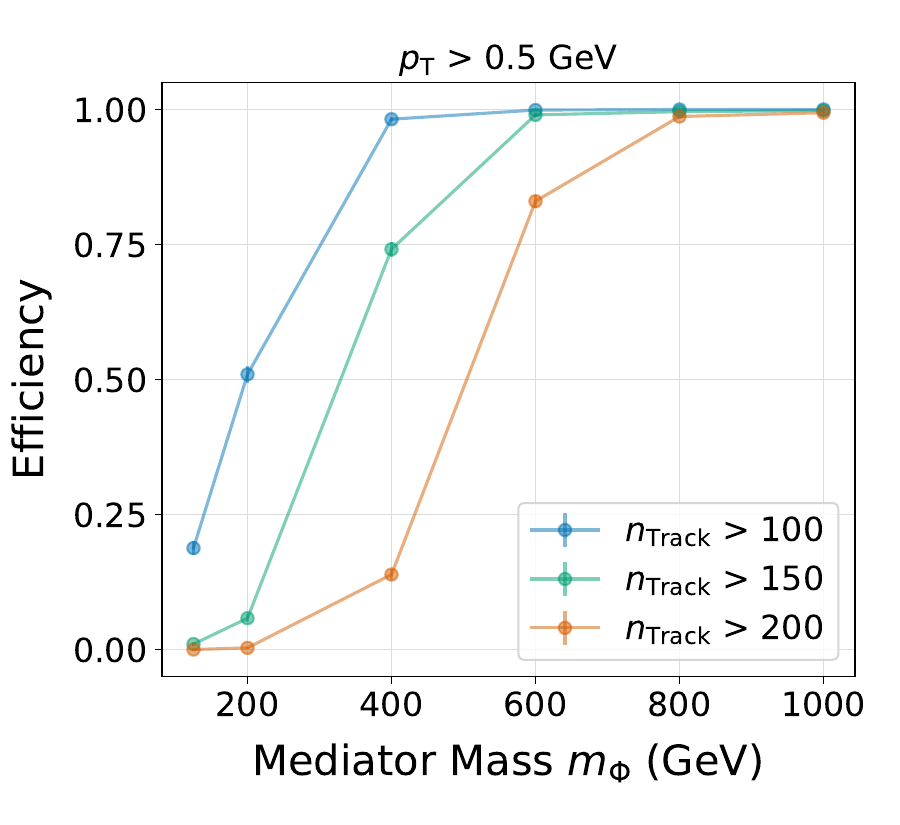}
         \caption{ }
         \label{subfig:suep_effs_05}
     \end{subfigure}
     \hfill
     \begin{subfigure}[b]{0.49\textwidth}
         \centering
        \includegraphics[width=\textwidth]{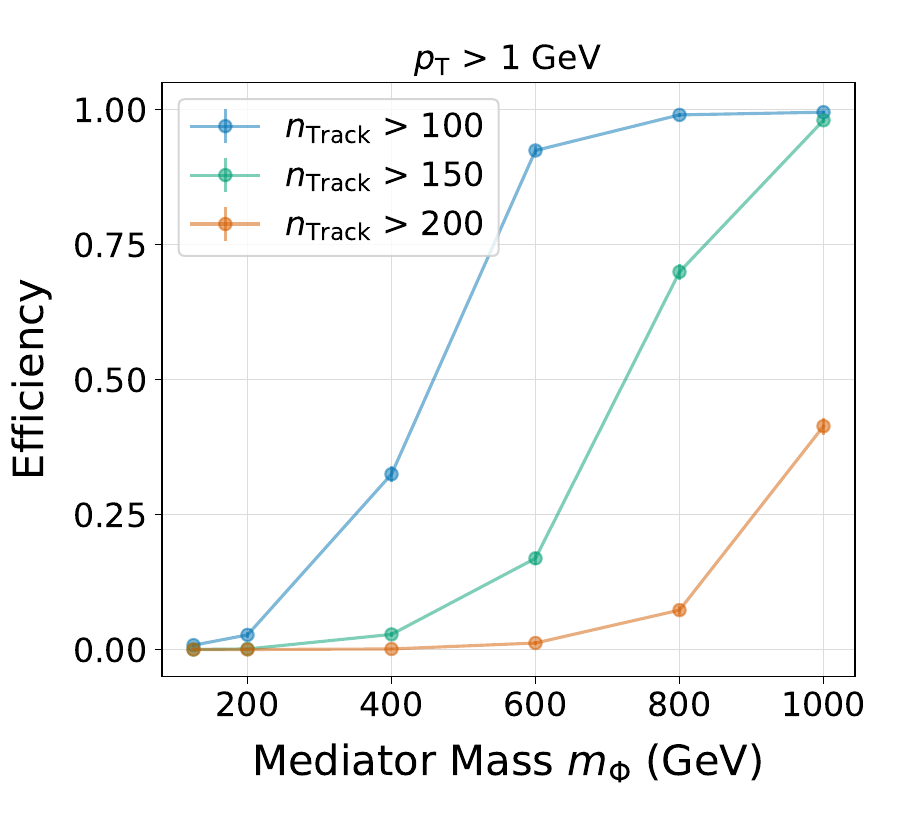}
         \caption{ }
         \label{subfig:suep_effs_1}
     \end{subfigure}
     
     \begin{subfigure}[b]{0.49\textwidth}
         \centering
        \includegraphics[width=\textwidth]{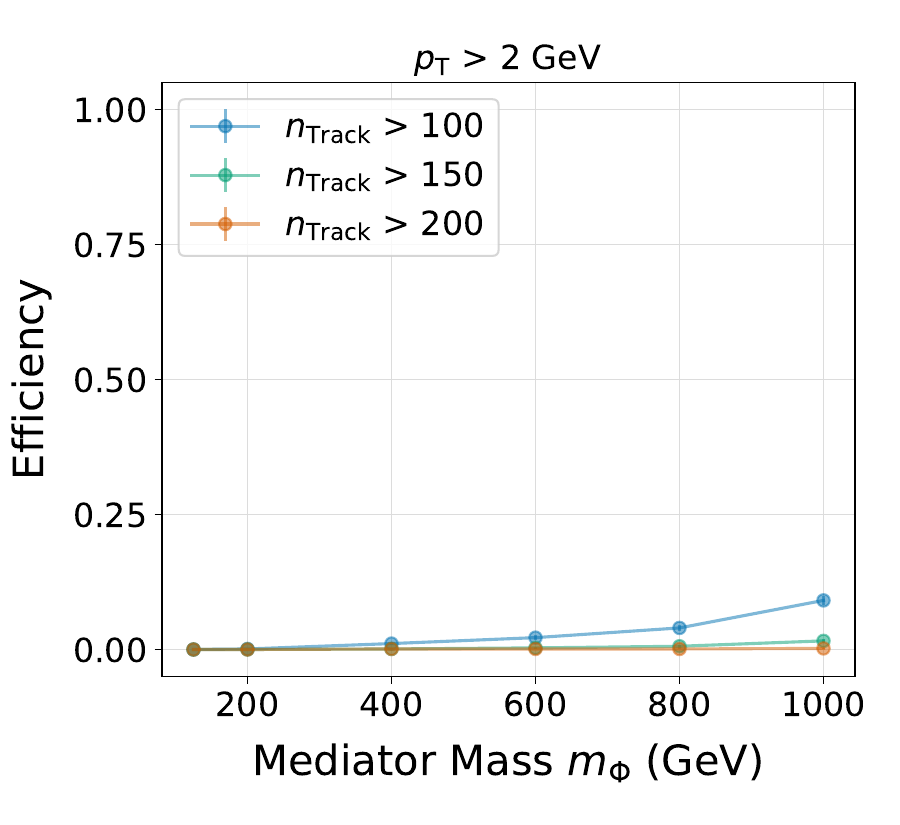}
         \caption{ }
         \label{subfig:suep_effs_2}
     \end{subfigure}     
    \caption{Efficiency, defined as number of events for which at least \nTrack within $\eta < 2.5$ pass the track reconstruction \pt threshold, as a function of mediator mass for minimum track \pt of 0.5 GeV (\subref{subfig:suep_effs_05}), 1 GeV (\subref{subfig:suep_effs_1}), and 2 GeV (\subref{subfig:suep_effs_2}). 
    }
    \label{fig:suep_effs}
\end{figure}

Efficiencies for all simulated mediator masses are shown as a function of minimum \nTrack for three different \pt thresholds in Figure~\ref{fig:suep_effs}. The efficiency increases with mediator mass: although the track \pt distribution does not depend on mediator mass, the larger number of tracks produced by heavy mediators means a sufficient number are reconstructed more frequently. With a track reconstruction threshold of $\pt > 0.5$~GeV, efficiencies of $50\%$ are obtained for mediators with masses above $200$~GeV with a $\nTrack > 100$ trigger, or for  mediators with masses above $400$~GeV with a $\nTrack > 150$ trigger. If the track reconstruction threshold increases to $\pt > 1$~GeV, $50\%$ efficiency is only possible for mediator masses above $\sim 600$ GeV. With a reconstruction threshold of $\pt > 2$~GeV, there is negligible efficiency for all signals with any \nTrack threshold. 

\begin{figure}[htb]
     \centering
     \begin{subfigure}[b]{0.49\textwidth}
         \centering
        \includegraphics[width=\textwidth]{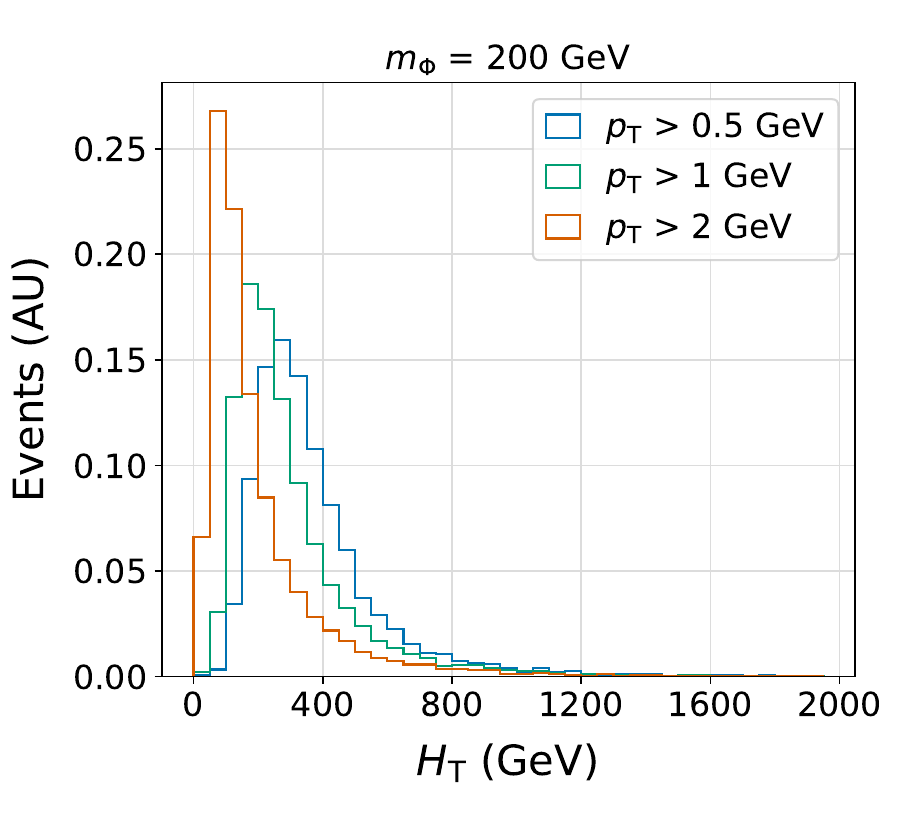}
         \caption{ }
         \label{subfig:suep_ht_200}
     \end{subfigure}
     \hfill
     \begin{subfigure}[b]{0.49\textwidth}
         \centering
        \includegraphics[width=\textwidth]{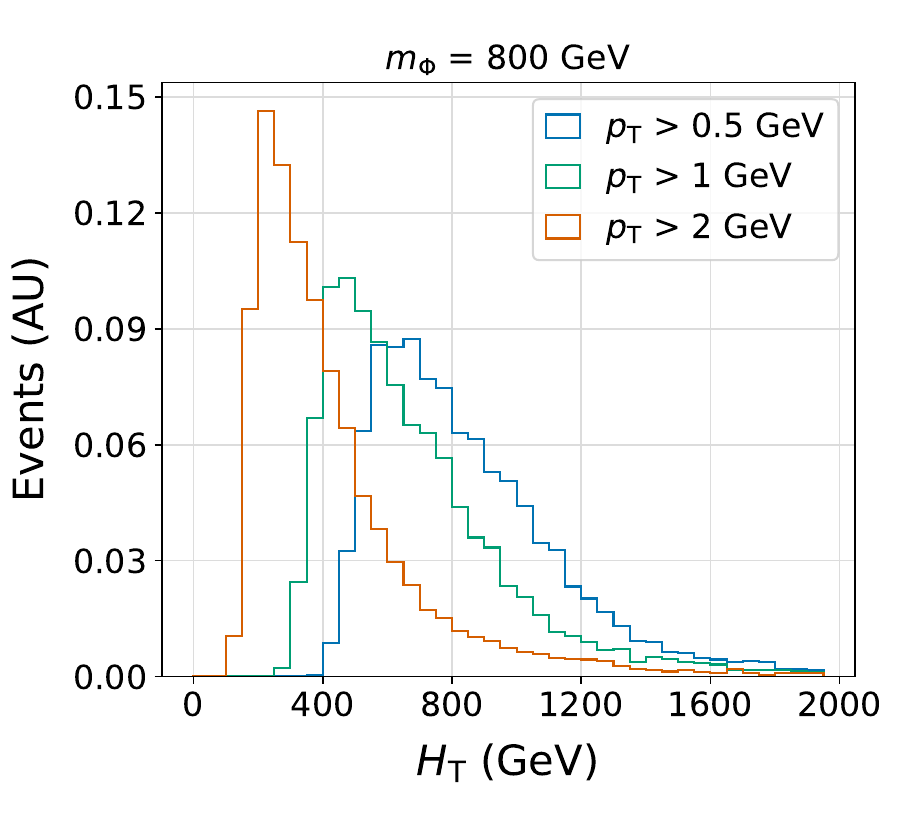}
         \caption{ }
         \label{subfig:suep_ht_800}
     \end{subfigure}
    \caption{\Ht of charged particles passing the specified track \pt threshold, for mediators of mass $200$~GeV (\subref{subfig:suep_ht_200}) and $800$~GeV (\subref{subfig:suep_ht_800}). 
    }
    \label{fig:suep_ht}
\end{figure}

For a track-trigger similar to the CMS HL-LHC upgrade, where $\pt > 2$~GeV is the baseline threshold, a trigger requiring a large track multiplicity is inefficient for most SUEP signal models. Other possible metrics for a SUEP trigger should be investigated, such as the scalar sum of jet energy, the scalar sum of track \pt (\Ht, shown in Figure~\ref{fig:suep_ht}), or event shapes such as isotropy or sphericity \cite{Cesarotti_2021, PhysRevD.20.2759}. At the LHC, SUEP final state particles are expected to be isotropically distributed in $\phi$, but localized in $\eta$, resulting in a ``belt of fire''. This characteristic shape has been shown to be an efficient discriminator at the HLT~\cite{Knapen_2017}. The most optimal trigger available to the LHC experiments will likely use \Ht or event shape in combination with the number of reconstructed tracks. 

%
\FloatBarrier
\subsection{Higgs portal to long-lived scalars}\label{sec:higgs_portal}

In this model, referred to here as the ``Higgs portal'', a SM Higgs boson decays to a pair of long-lived scalar particles, which each in turn decay to a pair of SM particles. Since the long-lived particle is not electromagnetically charged, no direct detection trigger strategies are available. The identifiable characteristic of this scenario is several relatively low momentum displaced jets. The low momentum nature of these jets presents a challenge for current L1 triggers because it is difficult for the calorimeter to distinguish these displaced jets from a large multi-jet background. Instead, this scenario can be targeted with a trigger that selects events with a number of displaced tracks.   

The benchmark model considered in this paper explores the well-motivated possibility that the SM Higgs boson acts as a mediator to a hidden sector via mixing with a new neutral scalar~\cite{ATLAS_displacedjets,Wells:2008xg,Gopalakrishna:2008dv}. A dark U(1) gauge group with a dark Higgs potential is introduced, leading to a total of two new particles, of which only the scalar has a mass low enough to be accessible at the LHC. This model serves as a good benchmark for a wide variety of Higgs portal scenarios, which will be a high priority target for the HL-LHC.  If the only available decay channel for the dark scalar is via its (small) mixing with the Higgs, this process will be suppressed, and the scalar will travel a macroscopic distance before decaying. This leads to the physical signature of displaced tracks arising from decay of the long-lived scalars into SM particles. These tracks generally have low \pt. The light mass of the scalar also means the tracks tend to have modest impact parameters, even when the scalar decays at a high radius. The branching ratio for the decay of the dark scalar follows that of the Higgs, restricted by its mass.

Higgs portal events are generated using MadGraph5\_aMC@NLO 2.9.3~\cite{Alwall:2014hca} and the SM + Dark Vector + Dark Higgs model described in Refs.~\cite{Curtin:2013fra,Curtin:2014cca}. The Higgs mixing parameter is taken to be $10^{-4}$ and the kinetic mixing parameter to be $10^{-10}$. The dark Z mass is set to 1 TeV, and effectively decoupled. In the generated scenario the dark scalar, $s$, is the only relevant new particle and it decays only into Standard Model particles. The properties of the dark scalar are varied across the mass range $m_s=5$, 8, 15, 25, 40, and 55 GeV and proper lifetime range 0.01, 0.1, and 1 ns. The scalar's branching ratio follows those of the Higgs, such that the scalars decay predominantly to $c\bar{c}$ and $\tau\bar{\tau}$ for $m_s = 5$ and 8 GeV and predominantly to $b\bar{b}$ for heavier $m_s$.

\begin{figure}[htbp]
\centering 
\includegraphics[width=.49\textwidth]{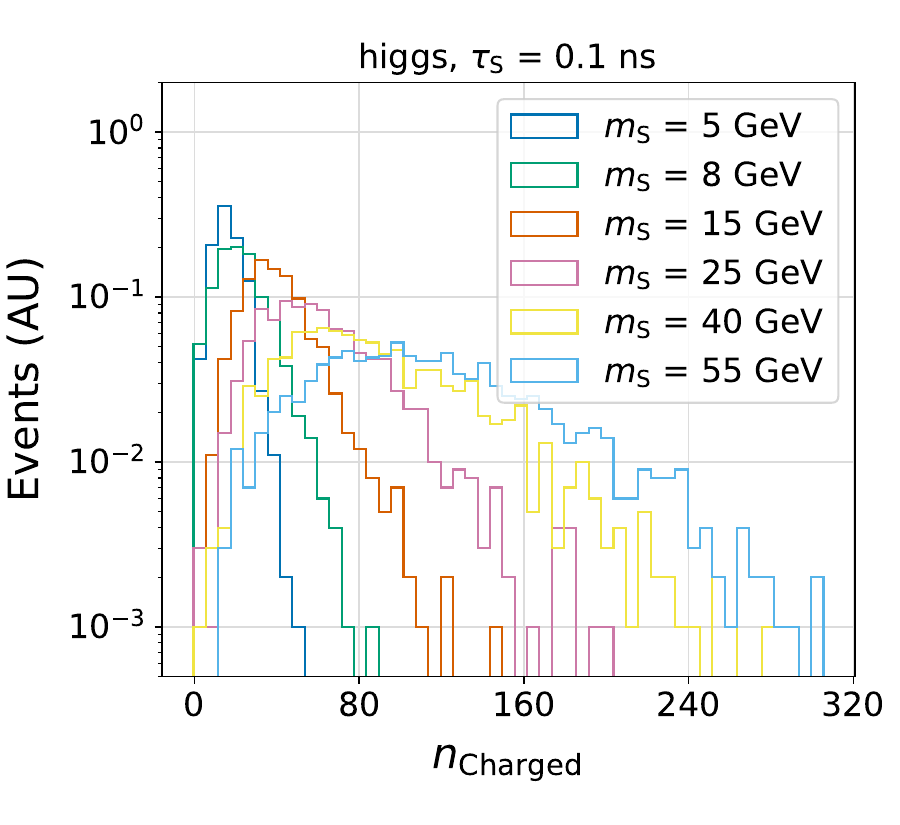}
\includegraphics[width=.49\textwidth]{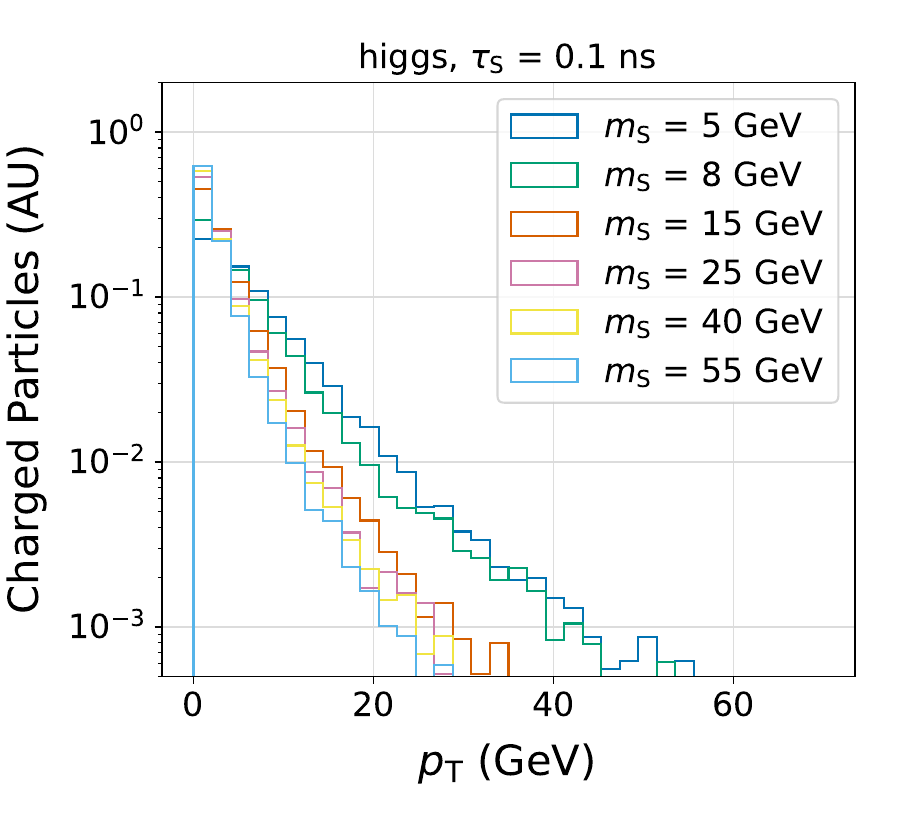}
\caption{\label{fig:higgs_nparticles_pt}
(Left) The number of displaced charged particles per event. Particles are required to be decay products of the long-lived scalar and have $\pt > 0.5$ GeV. (Right) Displaced charged particle \pt for each scalar mass. }
\end{figure}

The number of displaced charged particles per event and their \pt is shown in Figure~\ref{fig:higgs_nparticles_pt}. The low parent masses and fully hadronic final state result in a moderate number of low momentum particles. However, there are a sufficient number of displaced tracks to make a track-based trigger feasible.

A baseline track-level acceptance for this scenario is defined using requirements described in Table~\ref{tab:displaced_higgs_acc}. In addition to minimum \pt and $\eta$ requirements, displaced charged particles are required to be produced within a minimum radius and pass a minimum length requirement. These requirements ensure that displaced particles will traverse a sufficient number of tracker layers to be considered for reconstruction. 

\begin{table}[htbp]
\centering
\begin{tabular}{|c|c|}
\hline
Variable & Requirement \\
\hline
\multicolumn{2}{|c|}{Acceptance} \\
\hline
\Lxy & < 300 mm \\
$L_{\mathrm{track}}$ & > 200 mm \\
$|\dz|$ & $\geq 1 $ mm \\
\pt & > 0.5 GeV \\
$|\eta|$ & < 2.5 \\
\hline
\multicolumn{2}{|c|}{Efficiency} \\
\hline
\pt & > 0.5, 1, 2, 5, 10 GeV \\
$|\dz|$ & < 10, 20, 50, 100 mm \\
\hline 
\end{tabular}
\caption{\label{tab:displaced_higgs_acc} Details of the displaced track acceptance and efficiency definitions for the Higgs portal model. The \Lxy parameter refers to the maximum allowed production radius of the track, while the $L_{\mathrm{track}}$ variable refers the minimum track length.}
\end{table}

\begin{figure}[htbp]
\centering 
\includegraphics[width=0.48\textwidth]{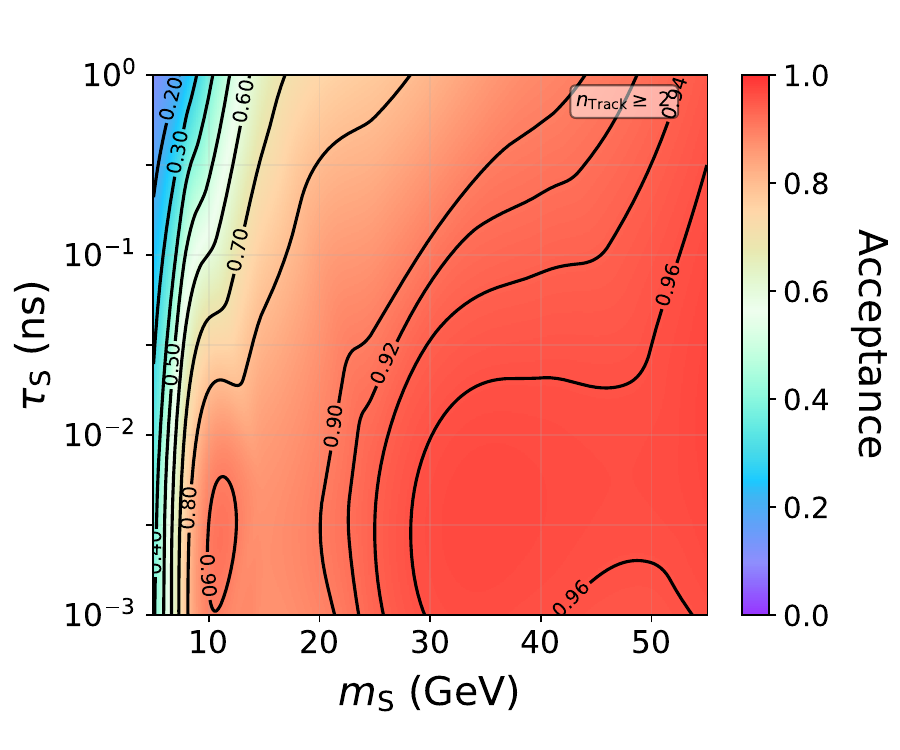} \hfill
\includegraphics[width=0.48\textwidth]{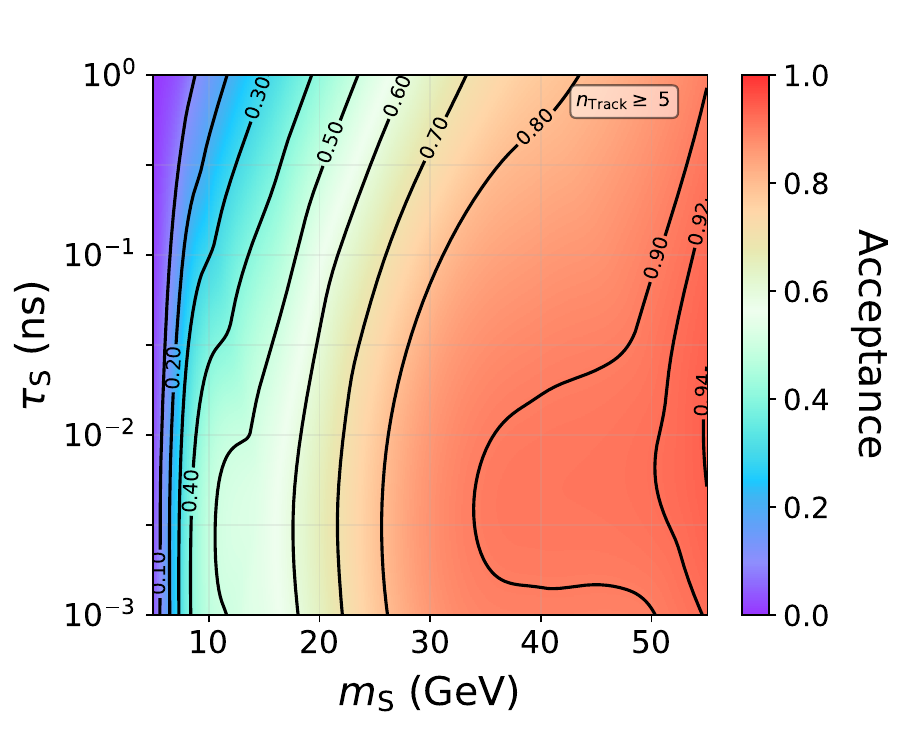} 
\caption{\label{fig:higgsacc} Acceptances for the Higgs portal scenario requiring at least two tracks (left) and least five tracks (right) passing the requirements listed in Table~\ref{tab:displaced_higgs_acc}.
}
\end{figure}

For this model, two event-level acceptance definitions are considered due to the low momenta of the displaced tracks. The first requires a minimum of five tracks per event to pass the acceptance criteria defined in Table~\ref{tab:displaced_higgs_acc}, and the second requires least two tracks per event. Event-level acceptances for these two definitions are shown in Figure~\ref{fig:higgsacc}.

Discontinuities in the event-level acceptances and efficiencies around scalar masses of $m_s=10$~GeV are a result of the sudden change in dominant decay modes, from pairs of charm quarks and tau leptons to $b\bar{b}$. The acceptance requirement has the largest impact at low masses, where there are fewer tracks likely to pass the minimum \pt threshold.

Track-level efficiency requirements corresponding to potential online tracking configurations are defined in Table~\ref{tab:displaced_higgs_acc}. The parameters considered include a minimum \pt and maximum \dz requirement, where the exact values are varied to understand the impact each selection has on the efficiency. Event-level efficiencies are defined as the fraction of events in acceptance which also have at least $\nTrack$ passing efficiency requirements.  

\begin{figure}[htbp]
\centering 
\includegraphics[width=0.49\textwidth]{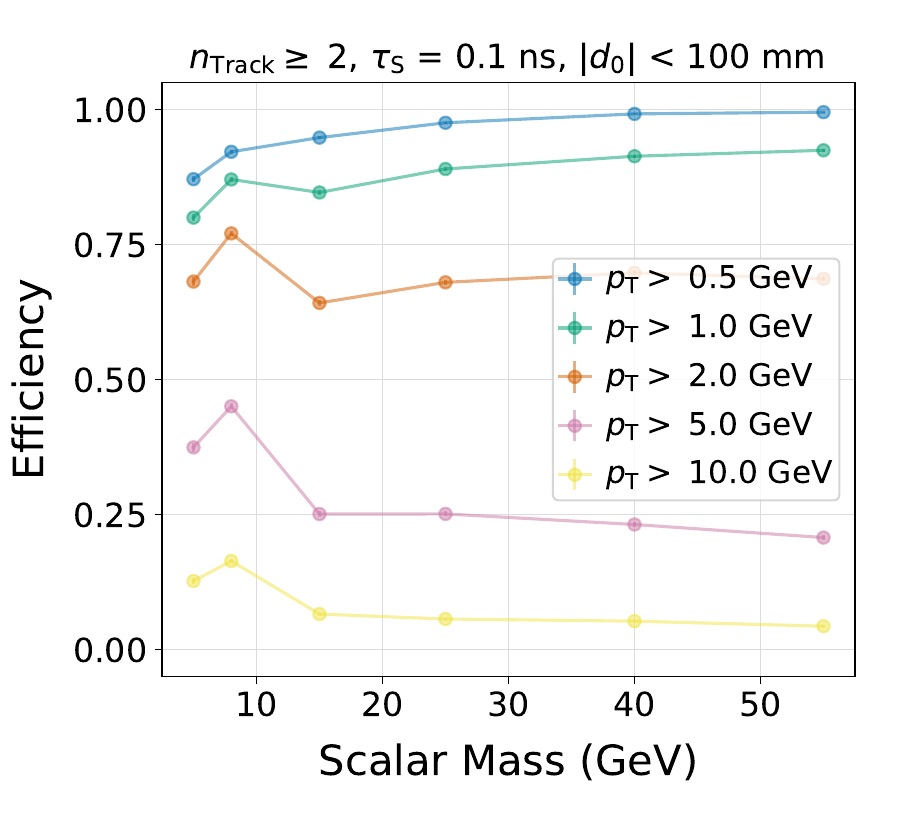}\hfill
\includegraphics[width=0.49\textwidth]{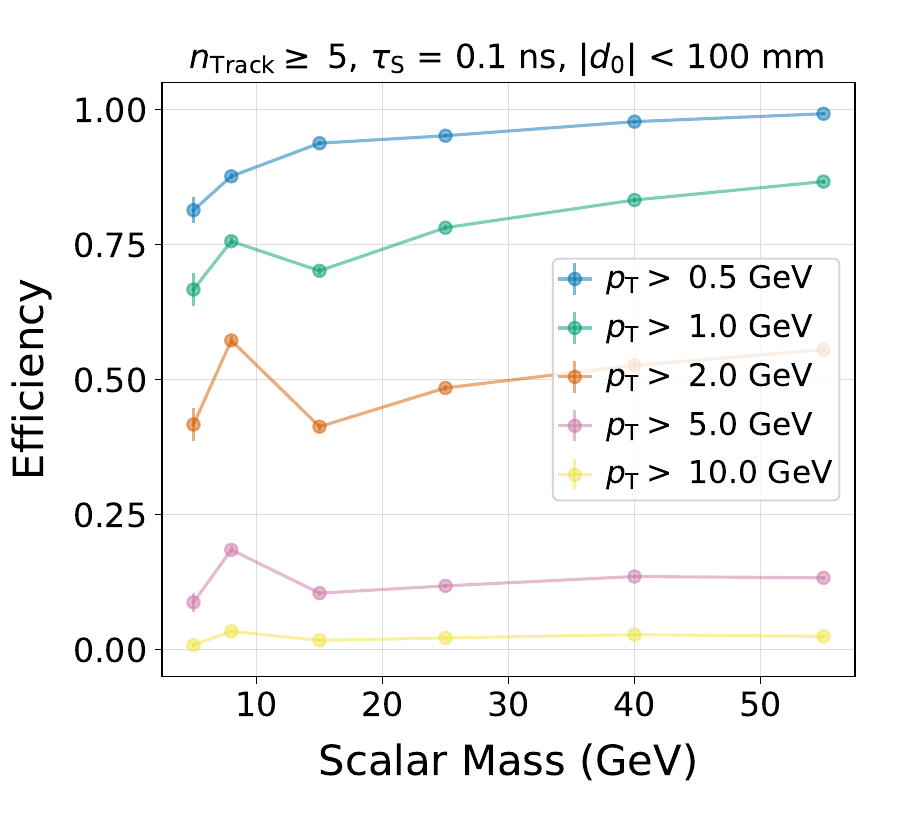}
\caption{\label{fig:vary_pt_higgs} Event-level efficiency as a function of scalar mass for a range of minimum track \pt values. Efficiencies are shown for the two (left) or five (right) scenarios. In all cases, the track reconstruction efficiency is assumed to decrease linearly with $\dz$ up until $|\dz|=100$~mm. 
}
\end{figure}

As with the acceptance definitions, two different values of $\nTrack$ are studied for the Higgs portal efficiencies. Efficiencies are defined on top of the corresponding \nTrack acceptance. Figures in this section largely correspond to $\nTrack \geq 5$, a likely realistic baseline for a trigger targeting hadronic long-lived particle decays. The additional study performed with $\nTrack \geq 2$ is to allow for direct comparison to the stau model. Results for this two-track trigger are used in Section~\ref{sec:comparisons} and directly compared in Figure~\ref{fig:vary_pt_higgs}.

As expected, the Higgs portal efficiencies are very sensitive to the minimum \pt requirement, as shown in Figure~\ref{fig:vary_pt_higgs}. The overall trend in efficiency is the same for each \nTrack considered. For a minimum \pt of 2 GeV, efficiencies for $\nTrack\geq5$ have dropped to around $50\%$ of their maximum value, and \pt thresholds any higher make this strategy ineffectual.

\begin{figure}[htbp]
\centering 
\includegraphics[width=0.49\textwidth]{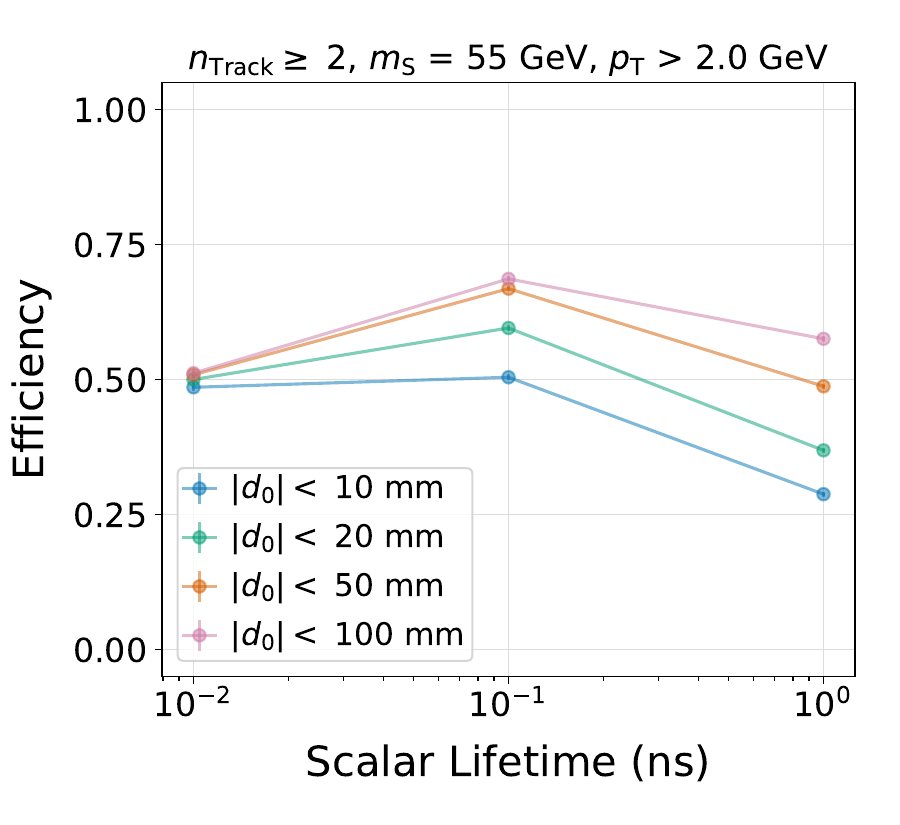} \hfill
\includegraphics[width=0.49\textwidth]{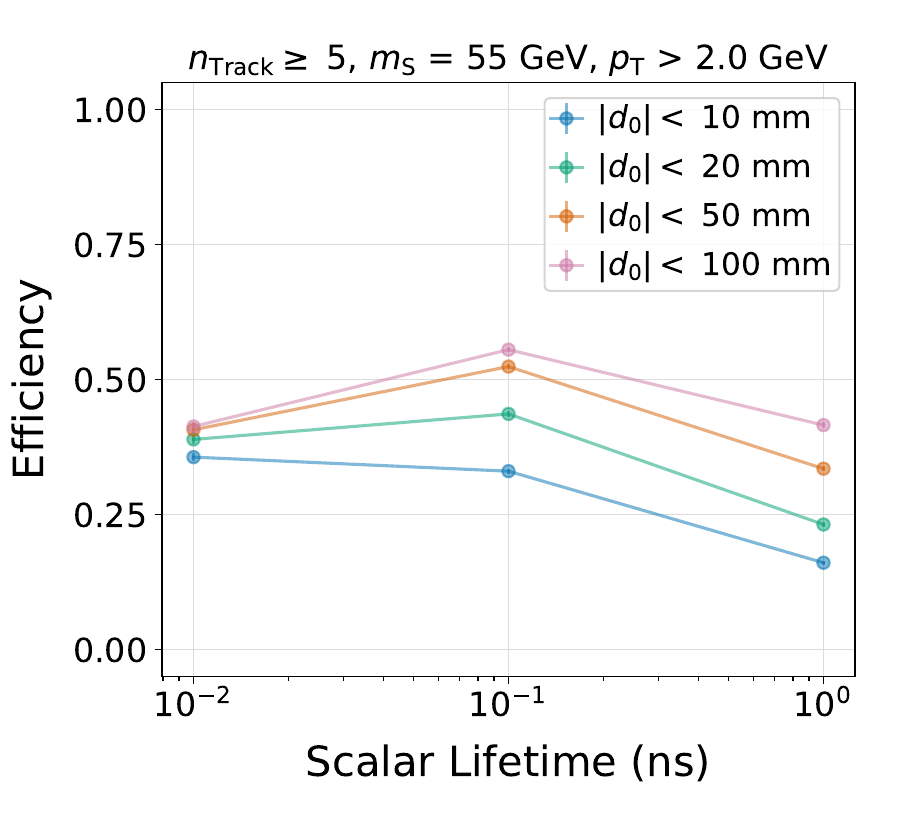}
\caption{\label{fig:vary_d0_higgs} Event-level efficiency as a function of long-lived scalar lifetime for different $|\dz|$ endpoints and \pt larger than 2 GeV. Efficiencies are shown for the two (left) or five (right) \nTrack scenarios. In all cases, the track reconstruction efficiency is assumed to decrease linearly with $\absdz$ up until the endpoint specified. 
}
\end{figure}

The Higgs portal efficiency's dependence on the \dz range, shown in Figure~\ref{fig:vary_d0_higgs} is more modest than the dependence on \pt. Endpoints of $|\dz|<1$~cm are at worst about $50\%$ as efficient as endpoints of $|\dz|<10$~cm for the lifetimes considered. With the high track multiplicity per event and long-lived particle boost expected in this model it is easier to find tracks with small \absdz, even for very displaced decays.

\begin{figure}[htbp]
\centering 
\includegraphics[width=0.48\textwidth]{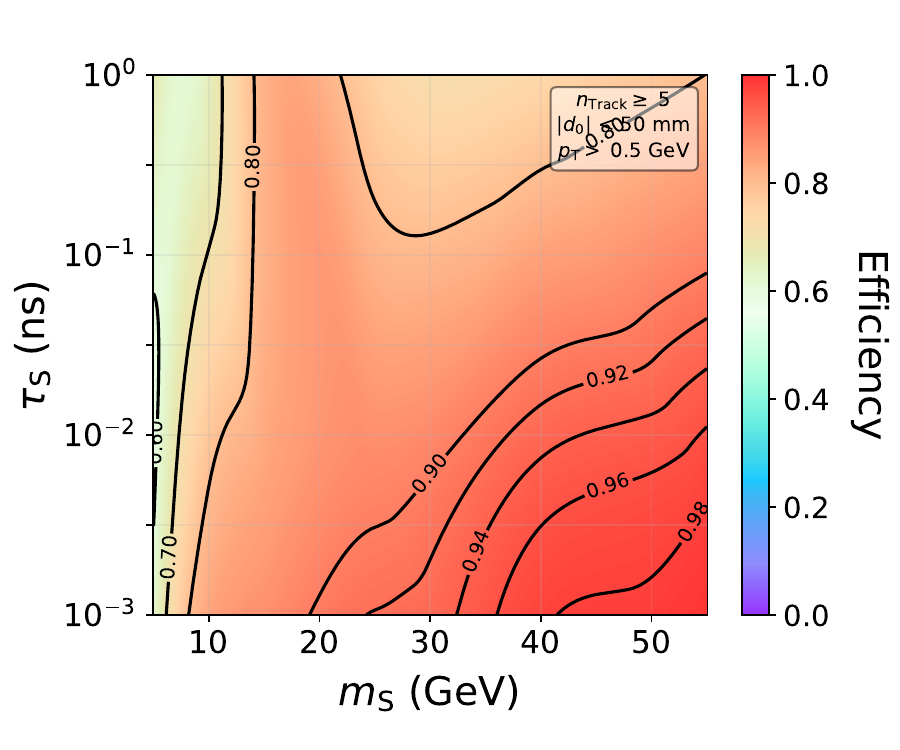} \hfill
\includegraphics[width=0.48\textwidth]{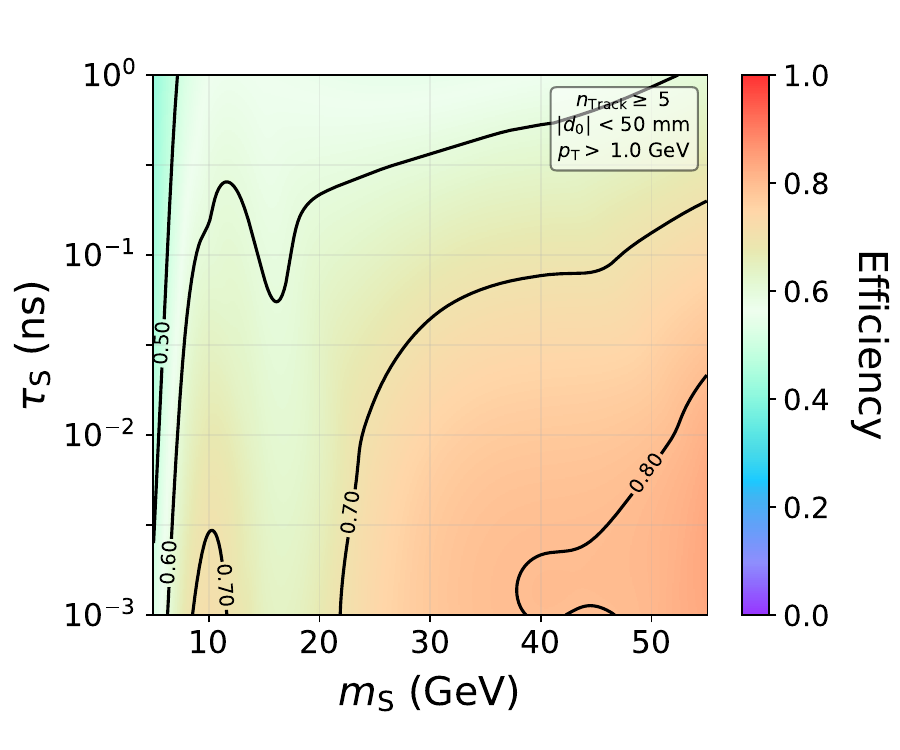} \\
\includegraphics[width=0.48\textwidth]{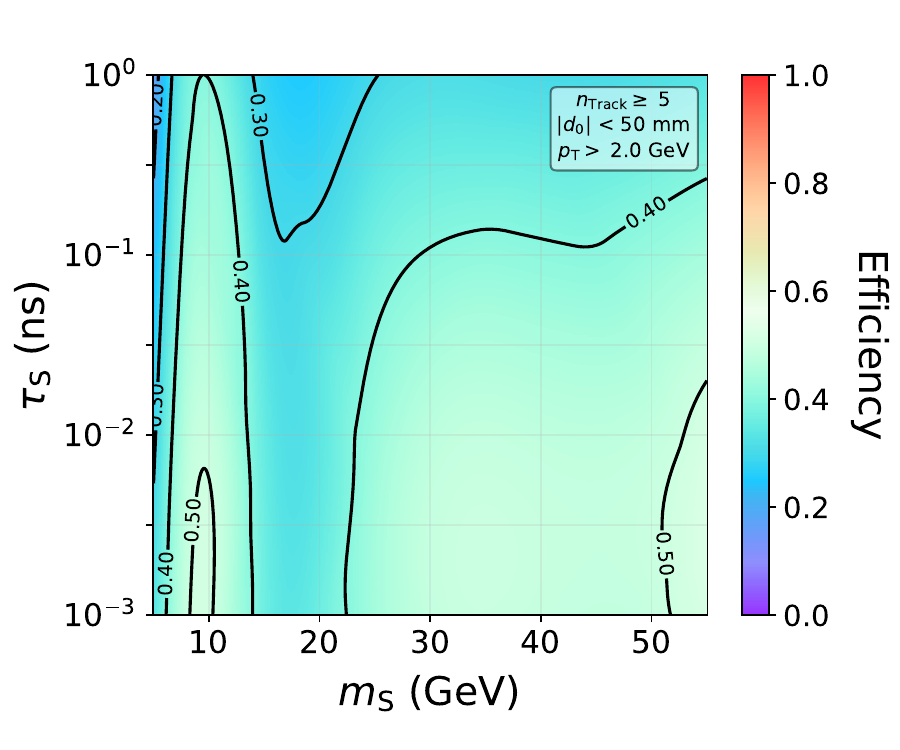} \hfill
\includegraphics[width=0.48\textwidth]{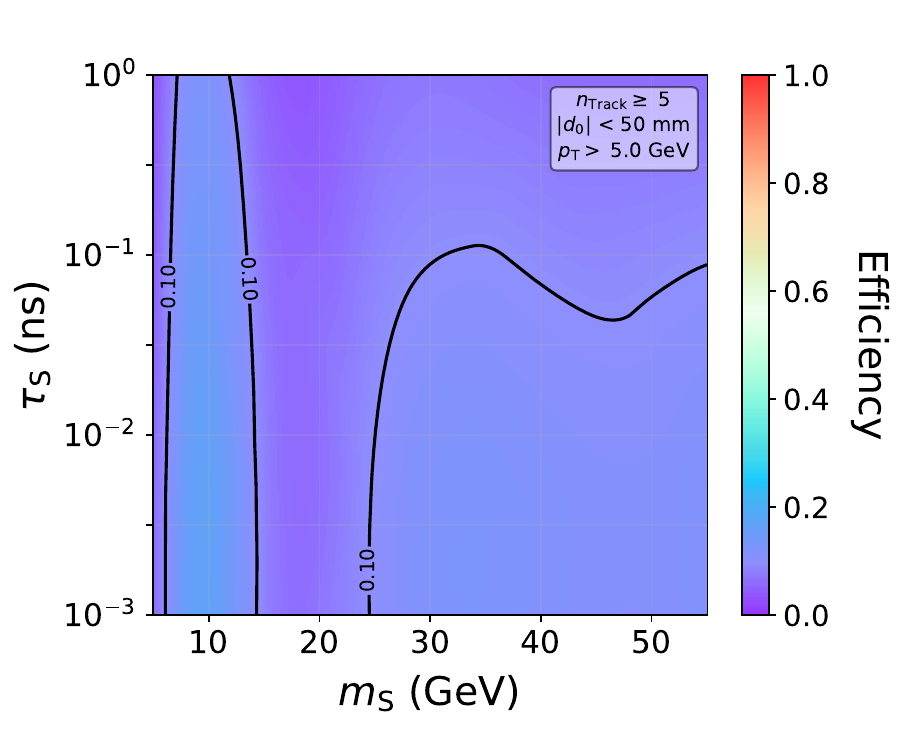}
\caption{\label{fig:higgs_2d} Efficiencies for the Higgs to long-lived scalar model for a variety of track \pt-thresholds:  0.5 GeV (top left), 1 GeV (top right), 2 GeV (bottom left), 5 GeV (bottom right). Across a wide range of masses and lifetimes, efficiency falls off steeply as the \pt threshold increases above 2 GeV. In all cases, a linearly parameterized \absdz efficiency is used with an endpoint at $50$~mm, and the five-track selection is used.
}
\end{figure}

Figure~\ref{fig:higgs_2d} shows the event-level efficiency as a function of scalar mass and lifetime for a variety of minimum \pt thresholds. The nominal threshold of $\pt>2$~GeV for the CMS Phase 2 tracker results in a peak efficiency of around $50\%$ for higher mass scalars. 

\FloatBarrier

\subsection{Long-lived staus}

Long-lived staus serve as a useful benchmark for two long-lived particle detection techniques. Staus can be either be directly detected as HSCPs or indirectly identified via displaced decay products. The current most comprehensive limits are from the OPAL Experiment, which exclude staus with mass $m_{\tilde{\tau}} < 90$~GeV for all possible lifetimes~\cite{OPAL:2005pwy}. LHC searches for for highly ionizing particles, disappearing tracks, and displaced leptons extend exclusions to larger stau masses~\cite{ATLAS:2022pib,CMS:2020atg,ATLAS:2022rme,ATLAS:2020wjh,CMS:2021kdm}. However, these searches do not cover the full lifetime space. Searches which aim to indirectly detect displaced decay products are limited by trigger \pt thresholds and efficiency for displaced particles. Searches which aim to directly detect charged particles such as the stau itself are currently trigger-limited when a track is not reconstructed in the muon system. 

The benchmark model considered here assumes direct production of a pair of staus in a gauge-mediated supersymmetry breaking (GMSB) scenario~\cite{evans-staus,Alwall:2008ag,LHCNewPhysicsWorkingGroup:2011mji}. While some GMSB slepton models take the three sleptons to be mass-degenerate, here the stau alone is taken to be the next-to-lightest supersymmetric particle (NLSP). The gravitino is the lightest supersymmetric particle in GMSB models. The stau decays to a tau and a gravitino via a small gravitational coupling, thereby gaining a significant lifetime.

Stau samples are generated and decayed with MadGraph5\_aMC@NLO 2.9.3. A simplified model is used wherein the left-handed and right-handed staus are assumed to be mass-degenerate and have a mixing angle $\sin \theta = 0.95$~\cite{ATLAS:2020wjh}. The gravitino is given a negligible mass of 1 GeV. Stau samples are generated for masses $m_{\tilde{\tau}}=100$, 200, 300, 400, 500, and $600$ GeV and lifetimes 0.001, 0.01, 0.1, 1 ns. For these lifetimes, displaced tracks from the decay of the tau (pions and light leptons) can be reconstructed. Additional samples are generated with 10 ns and infinite lifetimes. For these lifetimes, the stau is sufficiently long-lived that it can be directly identified as an anomalous prompt track with slow velocity or high ionization.

\subsubsection{Displaced leptons}

For lifetimes below $1$ ns, long-lived staus are most easily identified by their displaced decay products. In the GMSB model considered, where the light gravitino does not interact with the detector, the only visible signatures of the stau decays are the displaced taus. In the case of pair-produced staus, two taus are present in the final state, each the only visible product of its own displaced vertex. The taus then decay as usual, leading to a mix of final state scenarios ranging from fully leptonic when both taus decay leptonically to fully hadronic.

While relatively rare (12\% of di-tau events), the fully-leptonic scenario produces the distinctive signature of a pair of displaced, high-quality leptons, and is the easiest signature to target. The fully leptonic decay mode is the only scenario explored at the LHC thus far, with Run 2 results from both ATLAS~\cite{ATLAS:2020wjh} and CMS~\cite{CMS:2021kdm}. Triggering on leptonic tau decays is also the most straightforward. Standard photon triggers can be used to recover displaced electrons. Triggers which only require a track reconstructed in the muon system, and are agnostic to tracker activity or impact parameter requirements, are highly efficient for displaced muons up to $|\dz| \lesssim 10$~cm. In Run 2, displaced electrons and muons were both required to pass a relatively high \pt threshold, upwards of $40$~GeV. Because each lepton carries roughly 1/6 of the parent stau's energy, these thresholds are prohibitively high for the identification of low-mass staus. 

Hadronic decays of displaced taus are harder to identify. The complex algorithms that are used to identify prompt taus must be adapted to recover efficiency for taus with larger impact parameters. Approximately $15\%$ of taus decay to three charged particles. In this unique case, a visible displaced vertex can be reconstructed. The remaining $50\%$ of taus decay hadronically to a single charged particle. In both cases, charged particles may be accompanied by one or more neutral hadrons, and are always accompanied by a single tau neutrino. The fraction of the tau's momentum carried by the charged particles depends on the decay mode.  

\begin{figure}[htbp]
\centering 
\includegraphics[width=.49\textwidth]{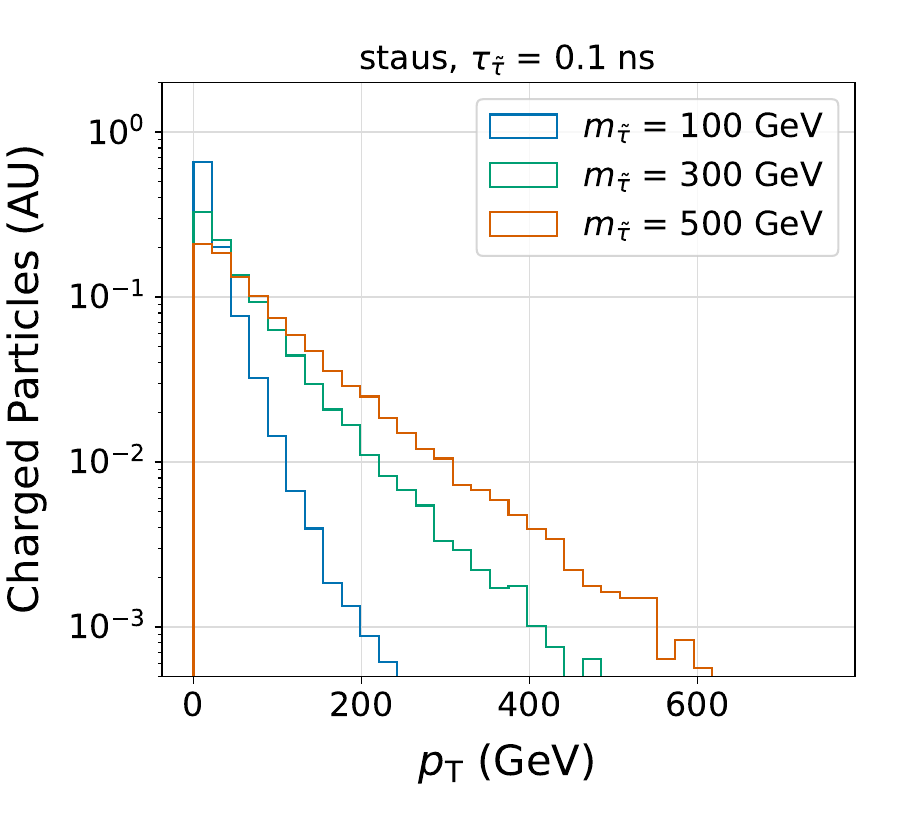}
\includegraphics[width=.49\textwidth]{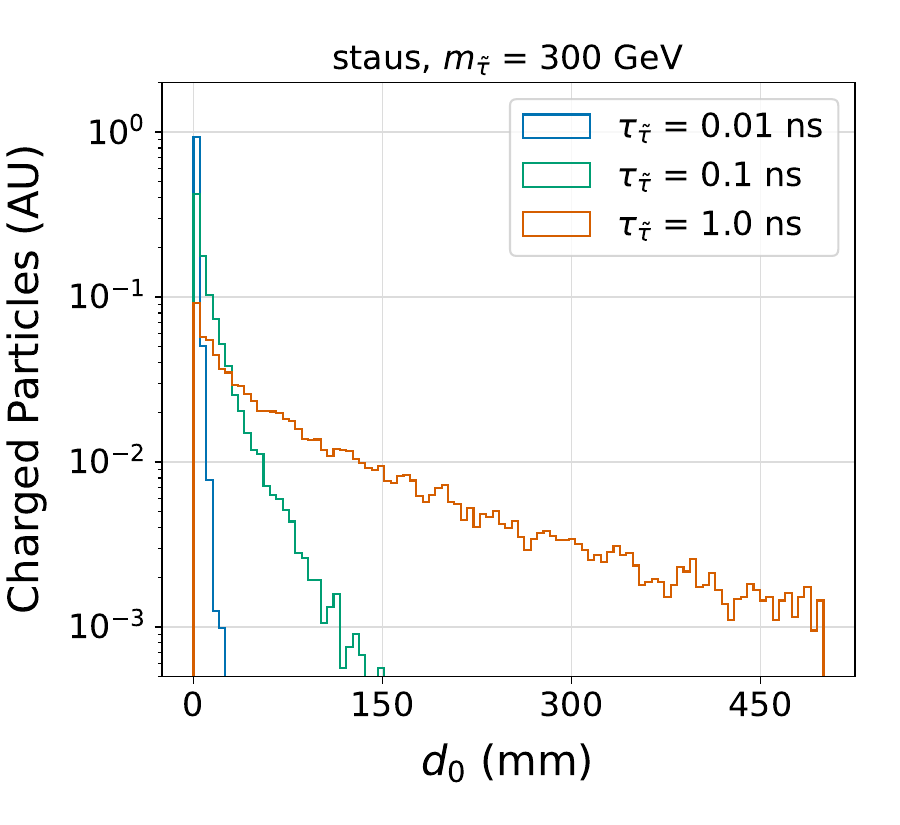}
\caption{\label{fig:displaced_tau_dists}
(Left) Displaced charged particle $\pt$. Particles are required to be decay products of the long-lived stau. (Right) Displaced charged particle \dz for varied lifetimes.}
\end{figure}

In all scenarios, at least two displaced tracks are produced per signal event. Figure~\ref{fig:displaced_tau_dists} shows the \pt and \absdz of displaced charged particles produced in the stau decay. For staus with masses above $100$~GeV, displaced tracks typically have \pt values well above the minimum thresholds for potential track-triggers. Displaced tracks tend to have larger impact parameters than in the Higgs portal model as a result of the larger parent mass. 

\begin{figure}[htbp]
\centering 
\includegraphics[width=.64\textwidth]{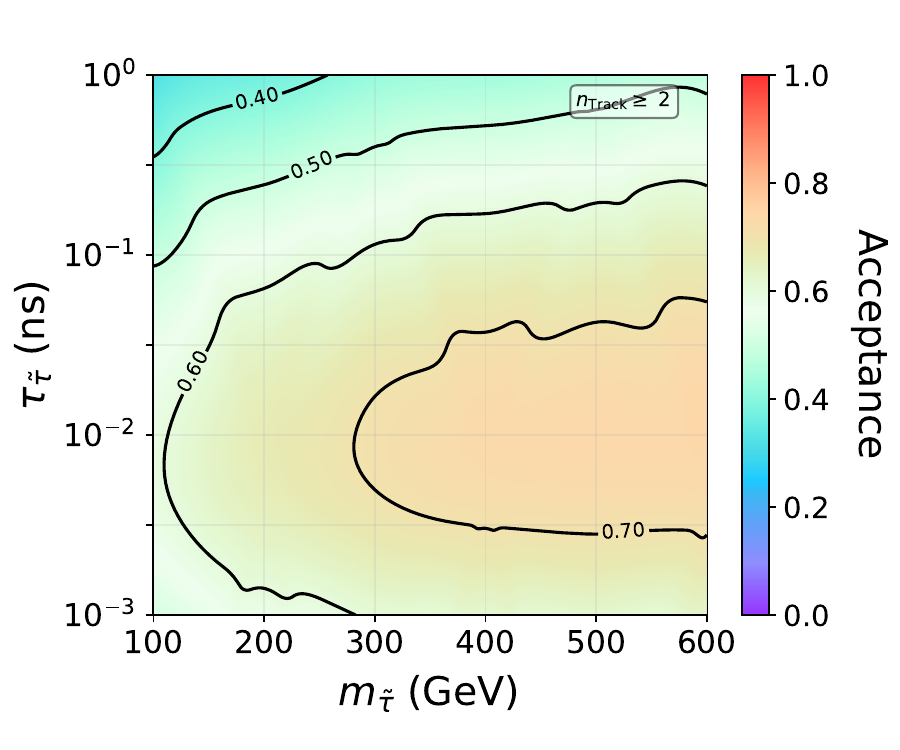}
\caption{\label{fig:displaced_lepton_acc} Acceptance for displaced lepton signature requiring at least two tracks and passing the requirements listed in Table~\ref{tab:displaced_higgs_acc}.
}
\end{figure}

This section follows a similar strategy to the Higgs portal model, as described in Section~\ref{sec:higgs_portal}. The baseline track acceptance is described in Table~\ref{tab:displaced_higgs_acc}. Events are considered to pass acceptance if they contain at least $2$~tracks passing all acceptance requirements. The resulting event-level acceptance, for a range of masses and lifetimes, can be seen in Figure~\ref{fig:displaced_lepton_acc}. 

\begin{figure}[htbp]
\centering 
\includegraphics[width=.64\textwidth]{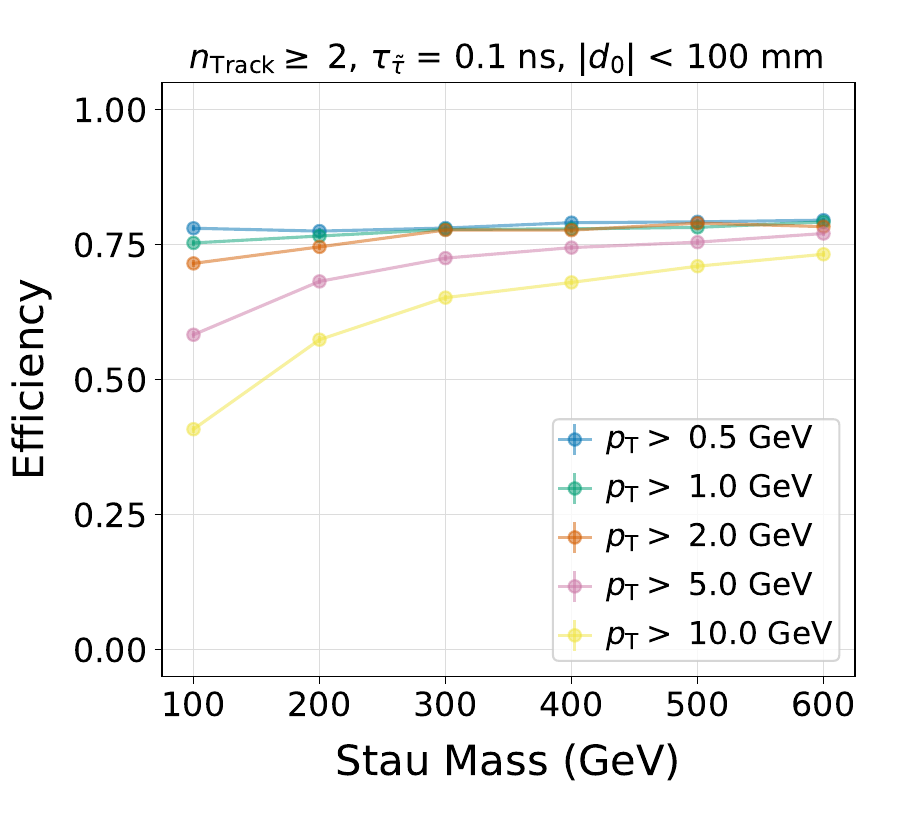}
\caption{\label{fig:vary_pt_staus} Efficiency as a function of stau mass for a range of minimum track \pt values. Events are defined to have passed if they contain two tracks with a \pt value larger than the minimum value, as well as a \absdz larger than 1 mm and smaller than 100 mm. The proper lifetime is $\tau=0.1$~ns. Track reconstruction efficiency decreases linearly from $100\%$ for prompt tracks to $0\%$ at $\absdz = 100$~mm.
}
\end{figure}

Efficiency requirements for individual tracks are also described in Table~\ref{tab:displaced_higgs_acc}. The event-level efficiency requires events have at least two tracks passing the per-track efficiency, with respect to events passing acceptance. A range of minimum \pt thresholds are explored, as shown in Figure~\ref{fig:vary_pt_staus}. In this scenario \pt thresholds have a minimal impact on overall efficiency, especially for higher stau masses.  For the CMS track-trigger baseline \pt threshold of 2 GeV, there is no significant reduction of efficiency compared to lower thresholds.

\begin{figure}[htbp]
\centering 
\includegraphics[width=0.49\textwidth]{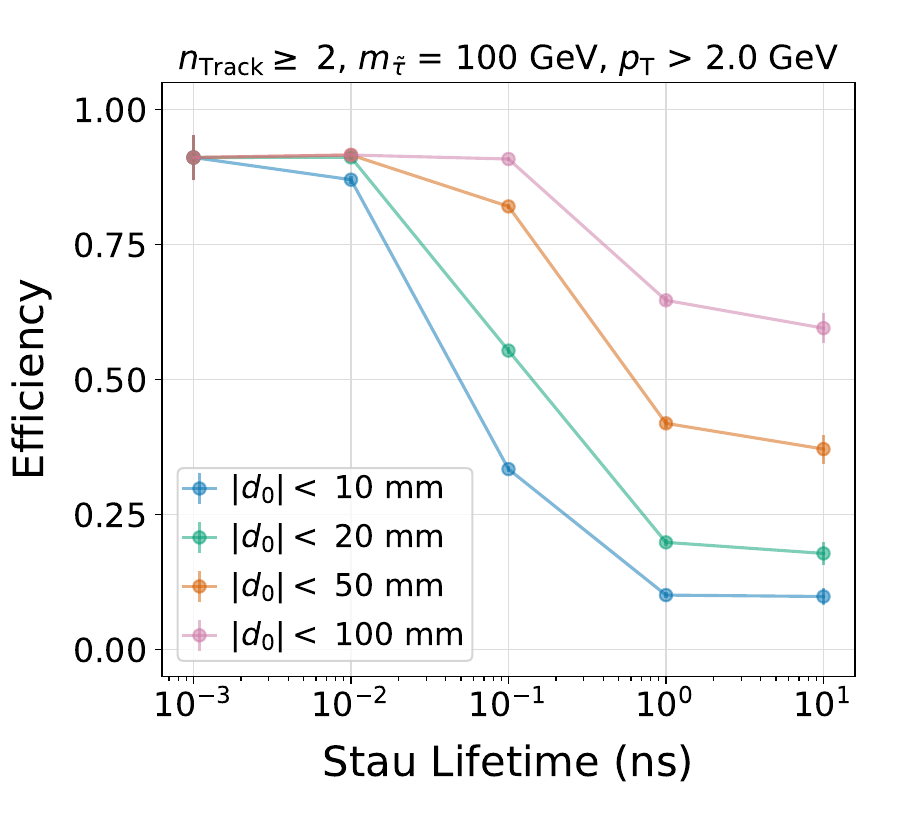}
\includegraphics[width=0.49\textwidth]{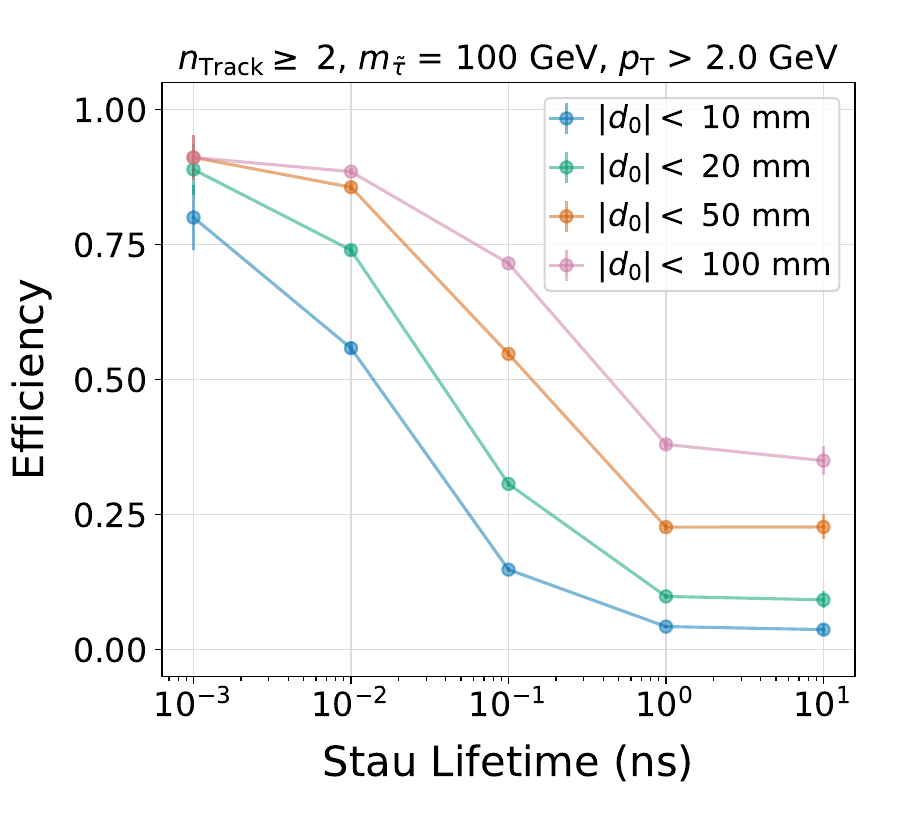} 
\caption{\label{fig:vary_slope_staus} Event-level efficiency as a function of stau lifetime for a $m_{\tilde{\tau}}=100$~GeV stau. (Left) Assumes $100\%$ efficiency up until a $\dz$-endpoint. (Right) Efficiency decreases linearly from $100\%$ for prompt tracks to $0\%$ at the $\dz$-endpoint. In both cases, a minimum track \pt of $2$~GeV and a minimum track \dz of $1$~mm are required. }
\end{figure}

The event-level efficiency for this model is much more sensitive to changes in the track-trigger efficiency as a function of \absdz. In this study, two parameterizations of efficiency are compared. The first parameterization is a binary model in which all tracks with \absdz less than a given value are assumed to be reconstructed. The second scenario follows the more realistic model described in Section~\ref{analysis-strat}. The efficiency for prompt tracks is assumed to be $100\%$, decreasing linearly to $0\%$ at the $|\dz|$ endpoint. The two scenarios are compared in Figure~\ref{fig:vary_slope_staus}, which shows event-level efficiencies for a range of stau lifetimes. Though the binary model results in a higher efficiency, by definition, the difference is modest ($<30\%$) for most lifetimes.  

\begin{figure}[htbp]
\centering 
\includegraphics[width=0.48\textwidth]{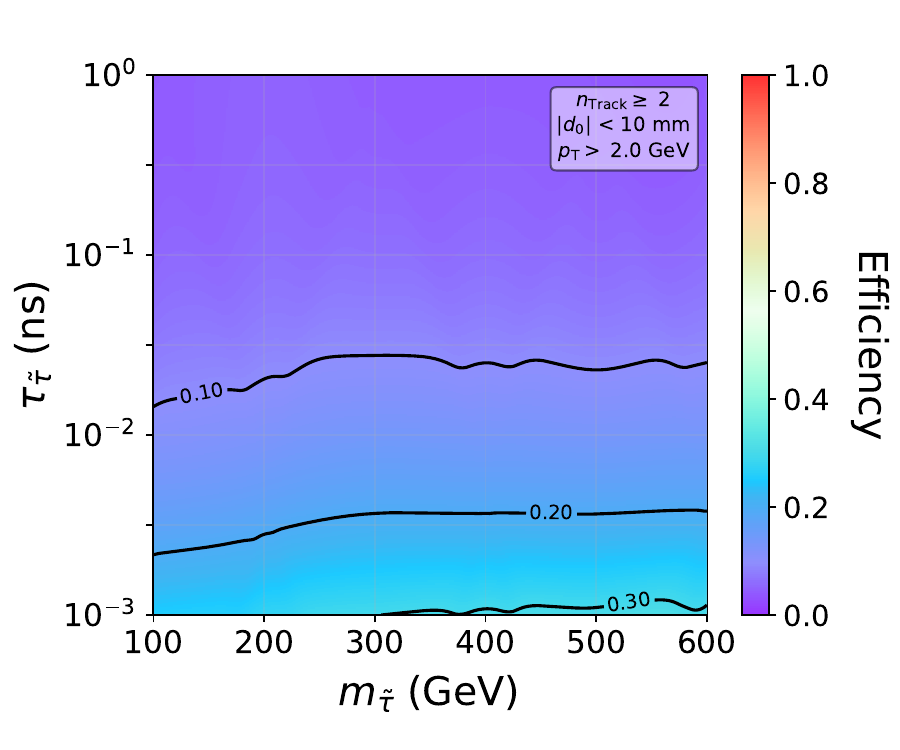}\hfill
\includegraphics[width=0.48\textwidth]{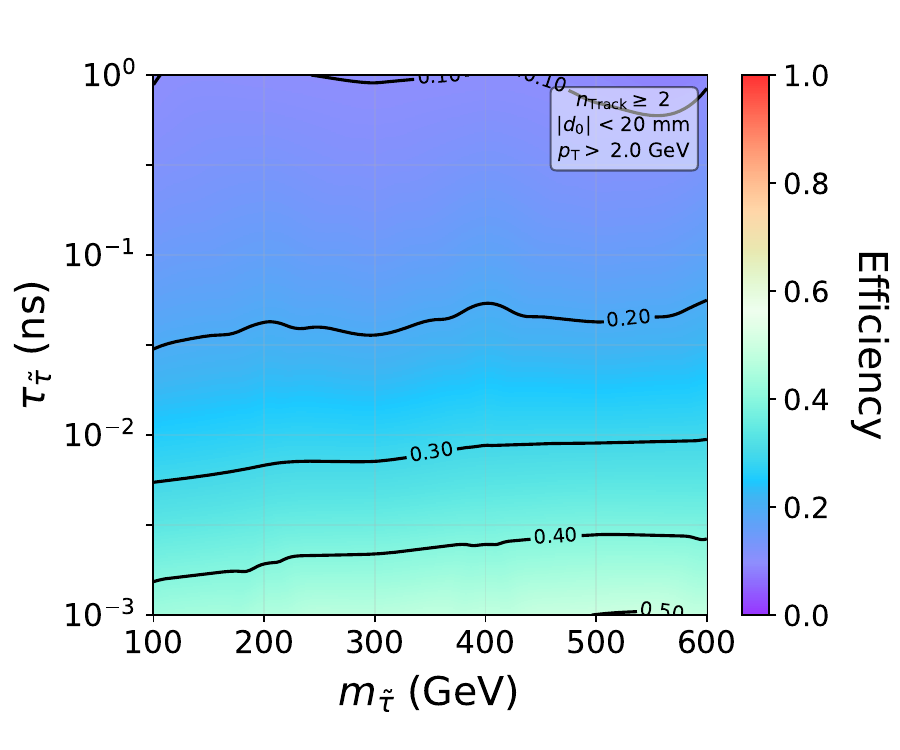} \\
\includegraphics[width=0.48\textwidth]{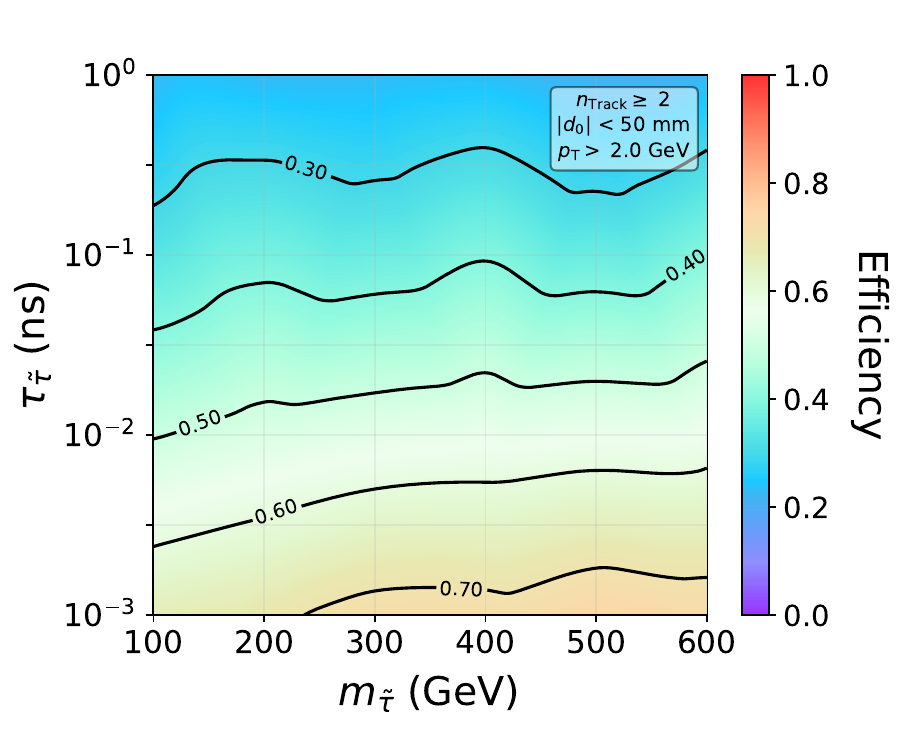}\hfill
\includegraphics[width=0.48\textwidth]{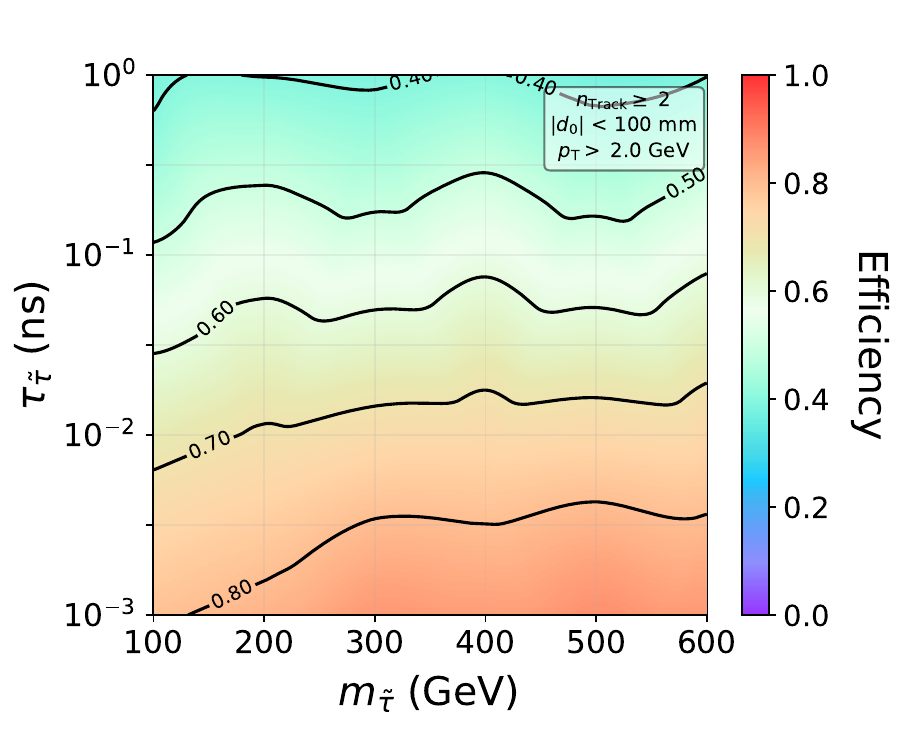}
\caption{\label{fig:staus_2d} Event-level efficiencies for the stau model for a variety of track $\dz$-endpoints:  $10$~mm (top left), $20$~mm (top right), $50$~mm (bottom left), $100$~mm (bottom right). In all cases, a linearly parameterized \dz efficiency is used. Tracks are required to have $\pt > 2$~GeV and events are required to have $\nTrack \geq 2$. 
}
\end{figure}

Figure~\ref{fig:staus_2d} shows the event-level efficiency as a function of stau mass and lifetime for a variety of \absdz-endpoints, assuming a linearly decreasing efficiency as a function of \absdz. For endpoints below $2$~cm, efficiencies are small ($\leq30\%$) across the plane, but for endpoints at $10$~cm, closer to typical offline efficiencies, efficiencies are high ($30-80\%$) for the full range of lifetimes. Given the low multiplicity of tracks expected per event with impact parameters roughly consistent with the stau decay radius, this sensitivity to \absdz-endpoint is expected. 

\FloatBarrier
\subsubsection{Direct detection}

For long-lived staus with proper lifetimes of $\tau \gtrsim 1$~ns, the charged stau will traverse a sufficient number of tracker layers such that its trajectory can be directly reconstructed as a prompt, high momentum, and isolated track. These staus will be slowly moving and highly ionizing with respect to Standard Model charged particles produced at the LHC. Most SM charged particles have mass $m<1$~GeV, $\beta \sim 1$, and can be approximated as minimum ionizing particles. In contrast, staus with masses $m \geq 100$~GeV will most often be produced with $\beta\gamma \sim 1$ and be slowly moving, $\beta<1$.

This study considers long-lived staus with masses between $100 \leq m_{\tilde{\tau}} \leq 1000$~GeV and proper lifetimes between $ 0.01 \leq \tau_{\tilde{\tau}} \leq 10$~ns. Two possible trigger strategies are studied. The first scenario looks for at least one high momentum and isolated track, and investigates the impact of geometric acceptance on trigger efficiency. This scenario is meant to be as inclusive as possible to direct long-lived particle detection, and can serve as a baseline for triggers which additionally incorporate delay or anomalous ionization measurements to reduce backgrounds. The second scenario assumes the outermost layer of the tracker also serves as a time of flight layer, and incorporates a timing measurement to further distinguish staus from SM tracks.

The selections considered for the inclusive trigger scenario are summarized in Table~\ref{tab:hscp_acc_eff}, and require at least one stau to decay within detector acceptance. The first geometric effect considered is the pseudorapidity range of the tracking detector or track-trigger. Typically the tracker is designed to emphasize efficiency and performance in the barrel. In comparison the endcap and further forward regions have degraded performance and higher background rates. The maximum pseudorapidity is varied from $|\eta| < 1.0, 2.5, 4.0$ to represent a barrel only scenario (if endcap rates are prohibitively high), a nominal scenario, and an extended scenario, respectively. 

Another geometric constraint is that the stau must traverse a sufficient number of tracker layers to be reconstructed. To investigate this effect, the minimum $\Lxy$ beyond which the stau must decay is varied  from $600 < \Lxy < 1200$~mm. The minimum \Lxy corresponds to particles which would traverse the pixel detector and at least three \pt-layers of the upgraded CMS tracker, or at least seven silicon layers in the upgraded ATLAS tracker. The maximum \Lxy roughly corresponds to the full tracker for both experiments. In all cases the minimum $|z|$ of the endcap is kept fixed at 3000 mm. 

At the HL-LHC, this inclusive trigger will likely include a track-based isolation requirement in order to reduce backgrounds. This study assumes sufficient tracker longitudinal impact parameter resolution such that any track-based stau isolation requirement would be fully efficient.

\begin{table}[htbp]
\centering
\begin{tabular}{|c|c|}
\hline
    Variable & Requirement  \\
    \hline
    \multicolumn{2}{|c|}{Acceptance}  \\ 
    \hline 
    $|\eta|$ &  $< 1.0, 2.5, 4.0$ \\
    $\Lxy$ &  $> 600,800,1000,1200$~mm \\
    OR $|z|$ & $>3000$~mm \\
    \hline
    \multicolumn{2}{|c|}{Inclusive Efficiency}  \\
    \hline
    \pt &  $> 10, 20, 50, 100$~GeV  \\ 
    \hline 
    \multicolumn{2}{|c|}{Time-of flight Efficiency}  \\
    \hline
    delay & $>0.25, 0.33, 0.50$~ns \\
    OR $\beta_{\mathrm{TOF}}$ & $<0.96, 0.95, 0.9$ \\
    OR $m_{\mathrm{TOF}}$ & $>15, 30, 60$~GeV \\ 
\hline
\end{tabular}
\caption{ \label{tab:hscp_acc_eff} Track acceptance and efficiency requirements considered for the inclusive high \pt charged particle trigger and for the time-of-flight trigger. Events are required to have at least one stau passing these selections. }
\end{table}

\begin{figure}[htbp]
\centering 
\includegraphics[width=.64\textwidth]{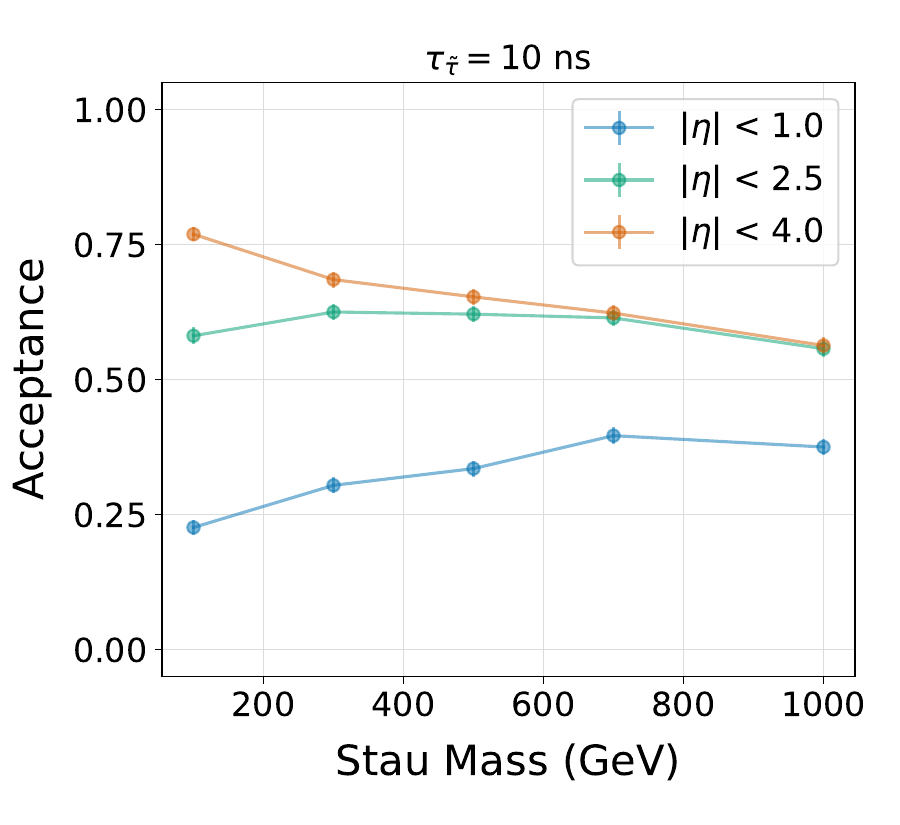}
\caption{\label{fig:hscp_isolatedtrack_eta} Stau model event-level acceptance for a variety of minimum tracker $|\eta|$ values. Acceptance is shown as a function of stau mass for proper lifetime $\tau=10$~ns. The stau is required decay beyond $\Lxy=1200$~mm.
}
\end{figure}

Figure~\ref{fig:hscp_isolatedtrack_eta} shows event-level acceptance as a function of varying the track-trigger maximum $|\eta|$ for a fixed $\Lxy=1200$~mm requirement. Varying the track trigger pseudorapidity has a larger effect on the event-level acceptance for lower mass staus. Extending the track trigger to $|\eta|<4.0$ would improve the acceptance for $m_{\tilde{\tau}} = 100$~GeV by $30\%$ but provides negligible benefit for $m_{\tilde{\tau}} = 1$~TeV. Extending the track trigger acceptance to the far forward region would be extremely challenging from a technical perspective, and offer little benefit in terms of physics reach for meta-stable charged particles because lower mass BSM particles also benefit from larger cross sections. In contrast, reducing the pseudorapidity coverage from $|\eta|<2.5$ to the barrel only scenario, $|\eta|<1.0$, would reduce the overall acceptance by approximately $50\%$ or more, and provides strong motivation for including the endcaps in any track trigger.

\begin{figure}[htb]
     \centering
     \begin{subfigure}[b]{0.49\textwidth}
         \centering
         \includegraphics[width=\textwidth]{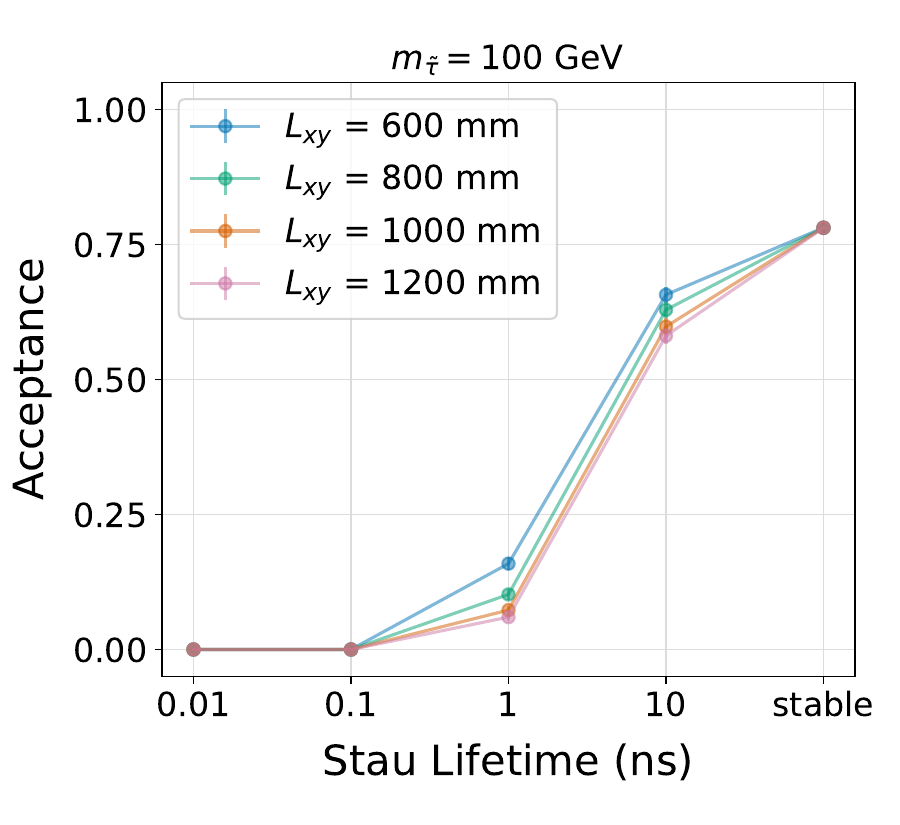}
         \caption{}
         \label{subfig:stau_m100_lxy}
     \end{subfigure}
     \hfill
     \begin{subfigure}[b]{0.49\textwidth}
         \centering
         \includegraphics[width=\textwidth]{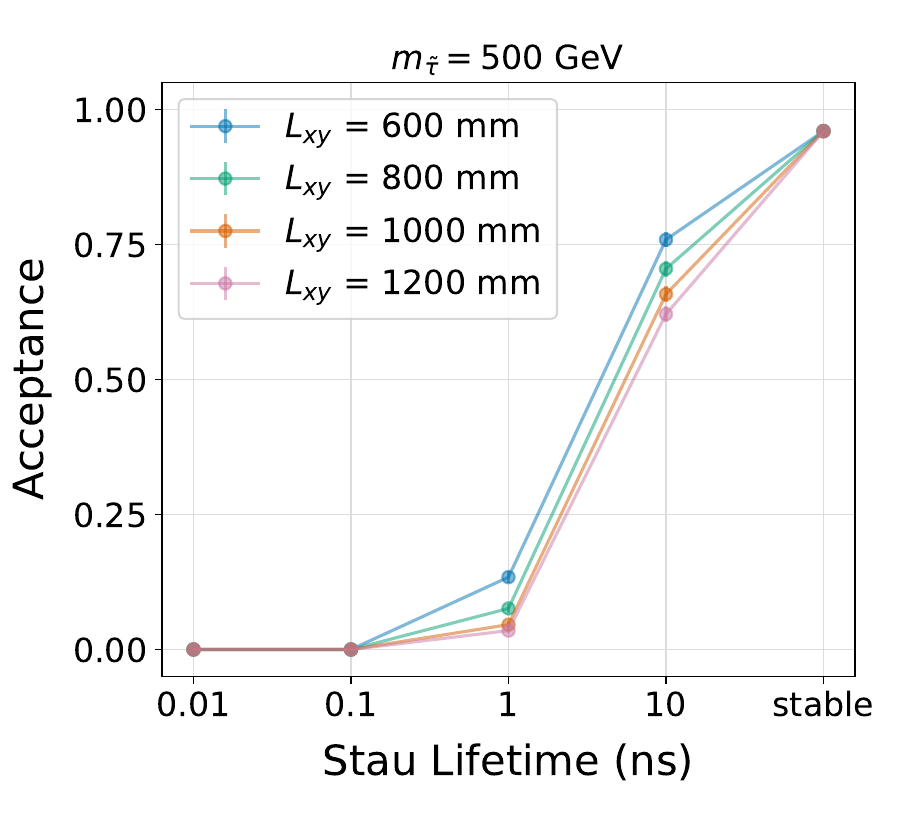}
         \caption{}
         \label{subfig:stau_m500_lxy}
     \end{subfigure}
    \caption{Stau event-level acceptance for a variety of minimum tracker $\Lxy$ values. Acceptance is shown as a function of stau lifetime for staus with mass $m_{\tilde{\tau}}=100$~GeV (\subref{subfig:stau_m100_lxy}) and $m_{\tilde{\tau}}=500$~GeV (\subref{subfig:stau_m500_lxy}). The track-trigger covers $|\eta|<2.5$. 
    }
    \label{fig:hscp_isolatedtrack_lxy}
\end{figure}

The impact of varying the minimum number of layers to reconstruct a track is shown in Figure~\ref{fig:hscp_isolatedtrack_lxy}. This study assumes a fixed pseudorapidity requirement of $|\eta|<2.5$. Reducing the minimum $\Lxy$ from $1200$~mm to $600$~mm, or requiring fewer layers per track, improves the overall acceptance for stau lifetimes of $\tau_{\tilde{\tau}} =1$~ns by roughly a factor of four. However, the event-level efficiency for this lifetime never exceeds that of a displaced track trigger, as shown in Figure~\ref{fig:vary_slope_staus}. Varying the number of layers required per track does not significantly improve the acceptance for lifetimes $\tau_{\tilde{\tau}} >1$~ns, because most staus decay well beyond the tracker, $\Lxy \sim 1.2$~m.

Increasing the minimum \pt required per track is a useful handle to reduce background rates while retaining nearly full signal efficiency. The impact of varying the minimum track \pt threshold on the event-level efficiency is shown in Figure~\ref{fig:hscp_isolatedtrack_pt} of Appendix~\ref{app:hscp}. All \pt thresholds considered are nearly fully efficient except for the lowest stau mass considered. 

The second track-trigger scenario investigates using a time-of-flight detector as an additional handle to improve signal-background discrimination. A timing layer similar to the planned CMS MIP timing detector is considered~\cite{Butler:2019rpu}. The timing layer is assumed to be located at roughly $\Lxy=1150$~mm, $z=3000$~mm, covering $|\eta|<2.5$, and provide a $50$~picosecond resolution timestamp for all charged particles passing through the detector. ATLAS also plans to incorporate precision timing at the HL-LHC, but only at larger pseudorapity~\cite{CERN-LHCC-2020-007}. For ATLAS, it may be more optimal to consider calorimeter or muon spectrometer timing measurements to target heavy meta-stable charged particles.

The time-of-flight detector's measurement for a given particle is computed according to
\begin{displaymath}
t_\mathrm{hit} = \frac{ L(p_{\mathrm{T}} , \eta)}{c \cdot p} \sqrt{p^2 + m^2}
\end{displaymath}
An increase in path length, $L$, due to the magnetic field is negligible for charged particles with $\pt > 10$~GeV in a $4$~Tesla solenoidal magnetic field, and therefore neglected by this study. The detector resolution is assumed to have a Gaussian uncertainty with a width of $\sigma=50$~ps. The spread of collisions in $z$ and the spread in collision time are assumed to follow a Gaussian distribution with widths $\sigma=50$~mm and $\sigma=200$~ps, respectively. This study assumes the hardware track-trigger is able to measure the track's longitudinal impact parameter, or origin in $z$, but not the primary vertex time. It is likely that computing the primary vertex time will only be possible in the HLT or offline, and that the uncertainty in the L1's time-of-flight measurement will be dominated by the beamspot's spread in time. For simplicity, this study focuses on stable lifetimes, and the selections considered are summarized in Table~\ref{tab:hscp_acc_eff}.

\begin{figure}[htb]
     \centering
     \begin{subfigure}[b]{0.49\textwidth}
         \centering
         \includegraphics[width=\textwidth]{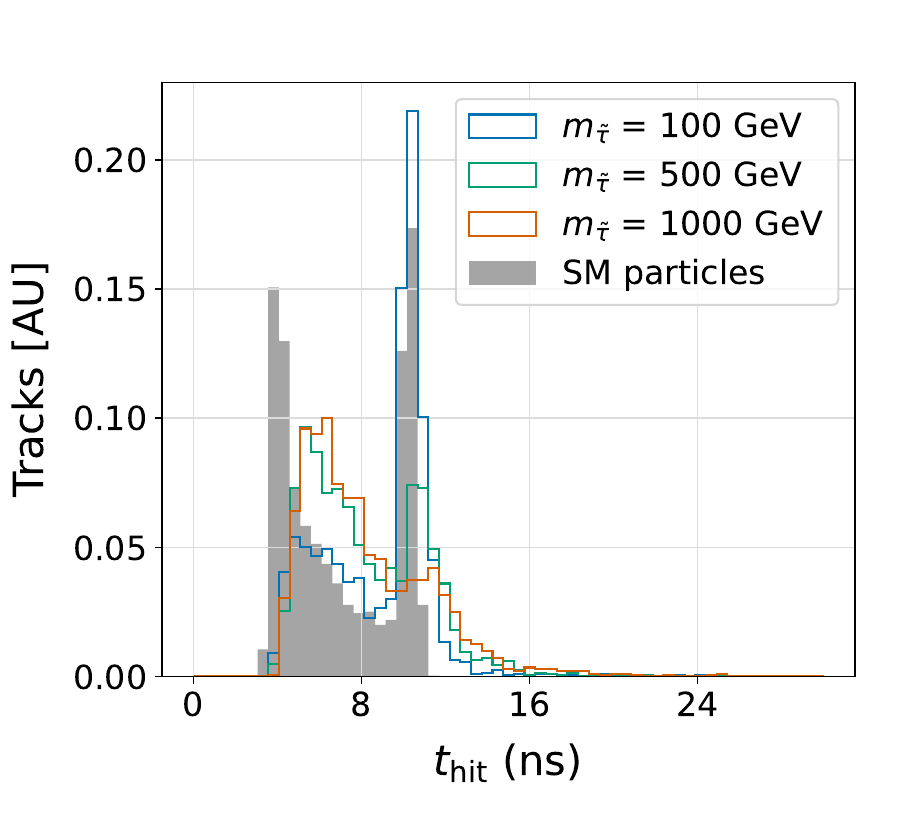}
         \caption{ }
         \label{subfig:hscp_tof}
     \end{subfigure}
     \hfill
     \begin{subfigure}[b]{0.49\textwidth}
         \centering
         \includegraphics[width=\textwidth]{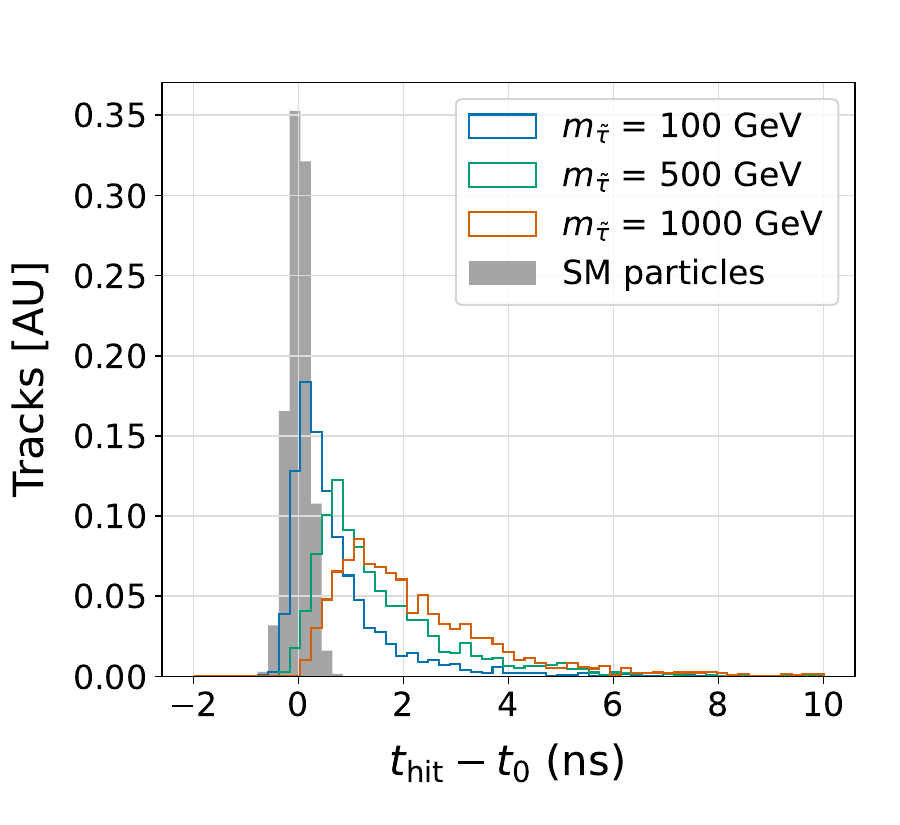}
         \caption{ }
         \label{subfig:hscp_delay}
     \end{subfigure}
    \caption{Timestamp as measured by the time-of-flight detector (\subref{subfig:hscp_tof}) and delay in time of arrival with respect to particles travelling at the speed of light (\subref{subfig:hscp_delay}). These distributions incorporate a realistic spread in collision time and $z$ position, as well as the timing layer's temporal resolution. SM particles with $\pt>10$~GeV from the signal samples are shown for comparison.
    }
    \label{fig:hscp_delay}
\end{figure}

Time-of-flight and delay distributions are shown in Figure~\ref{fig:hscp_delay} for different stau masses. For comparison, a ``background'' sample is simulated by using SM particles with $\pt > 10$~GeV in the signal samples as a proxy. The time-of-flight measurements contain two peaks, at $t_{\mathrm{hit}}\sim 3$ and $t_{\mathrm{hit}}\sim 10$~ns, corresponding to the difference in path length for tracks which traverse the barrel or the endcaps, respectively. The delay is defined to be the time-of-flight measurement subtracted by the time it would take a particle traveling at the speed of light to reach the same region of the detector. Staus arrive noticeably later than background particles, and the measured delay in arrival increases with stau mass. 

\begin{figure}[htb]
     \centering
     \begin{subfigure}[b]{0.49\textwidth}
         \centering
         \includegraphics[width=\textwidth]{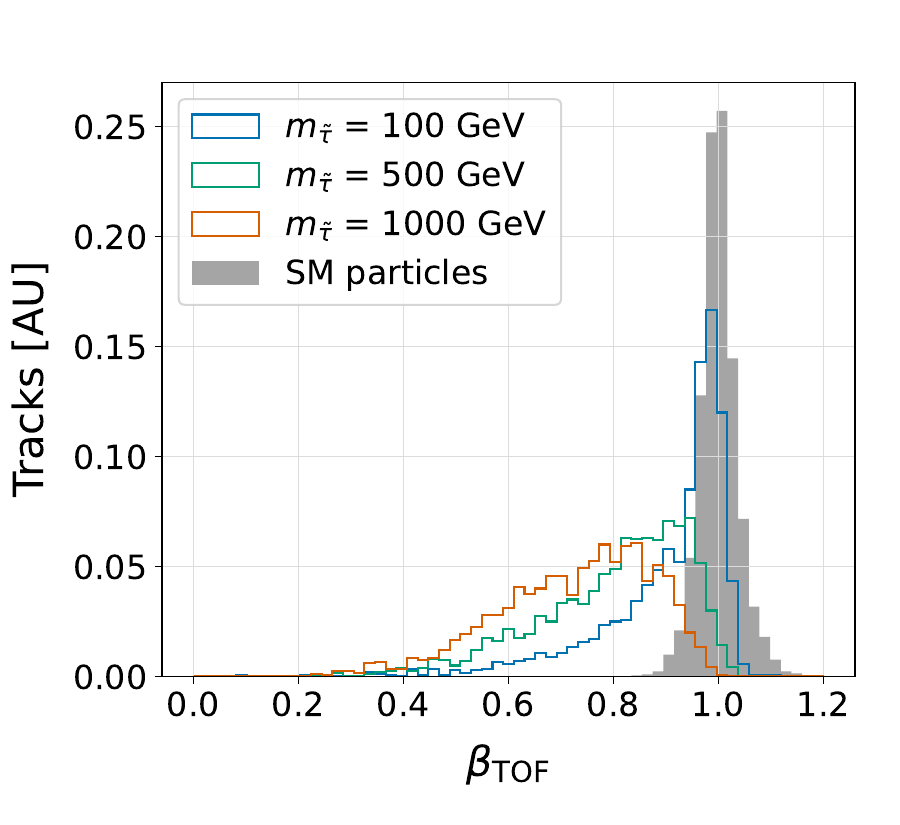}
         \caption{ }
         \label{subfig:hscp_beta}
     \end{subfigure}
     \hfill
     \begin{subfigure}[b]{0.49\textwidth}
         \centering
         \includegraphics[width=\textwidth]{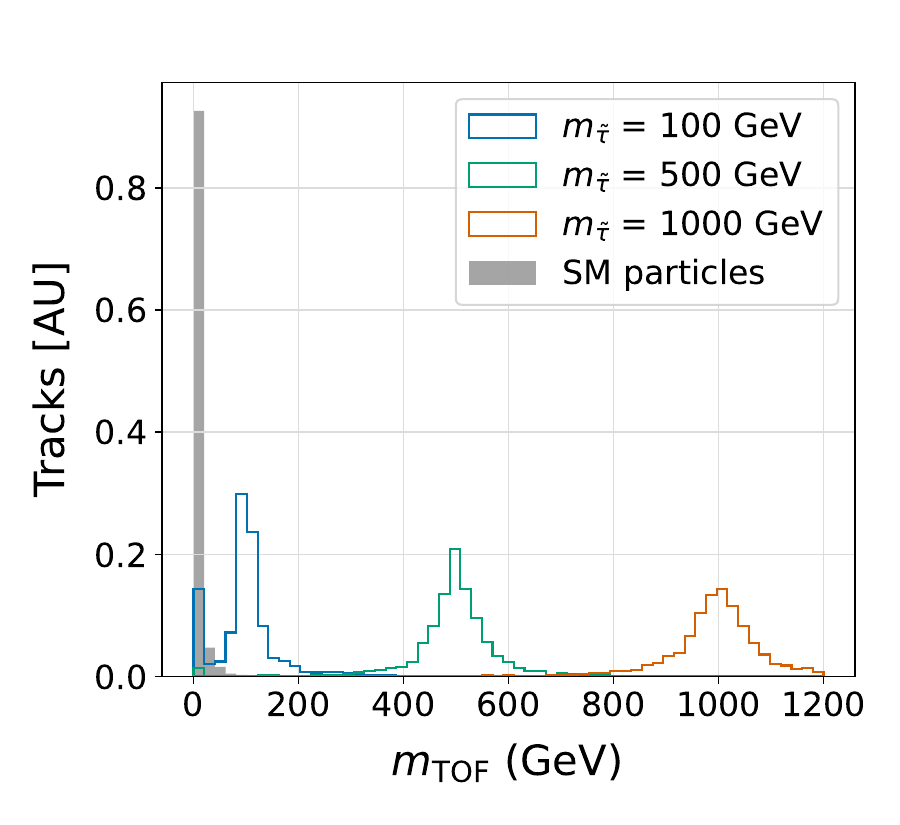}
         \caption{ }
         \label{subfig:hscp_mass}
     \end{subfigure}
    \caption{Measured $\beta$ (\subref{subfig:hscp_beta}) and mass  (\subref{subfig:hscp_mass}) for different stau masses. These distributions incorporate a realistic spread in collision time and $z$ position, as well as the timing layer's temporal resolution. SM particles with $\pt>10$~GeV from the signal samples are shown for comparison.
    }
    \label{fig:hscp_beta_mass}
\end{figure}

With the time-of-flight measurement it is also possible to extract the stau velocity, $\beta$, and mass as shown in Figure~\ref{fig:hscp_beta_mass}. Computing $\beta_{\mathrm{TOF}}$ accounts for differences in path length, due to the stau $\eta$ and the longitudinal position of the primary vertex $z$. As expected, background tracks peak at $\beta_{\mathrm{TOF}}=1$, while the measured velocity decreases with increasing mass.

Computing a charged particle's mass requires a track momentum measurement, and the mass resolution can be described as 
\begin{displaymath}
(\Delta m_{\mathrm{TOF}})^2 = m^2 \Bigg[ \Big( \frac{ \Delta  p_{\mathrm{T}} }{p_{\mathrm{T}}} \Big)^2 + \Big(\frac{1}{1-\beta^2} \Big)^2 \Big( \frac{\sigma_{t_{\mathrm{hit}}}}{t_{\mathrm{hit}}} \Big)^2 \Bigg]  
\end{displaymath}
For simplicity, this study assumes a \pt resolution of $1\%$ which increases up to $10\%$ for particles with \pt of $1$~TeV. This assumption is somewhat optimistic for the endcaps, but reasonable for the barrel~\cite{CERN-LHCC-2017-009,CERN-LHCC-2020-004}. Measured masses peak at the actual stau mass, while background is peaked near zero, with long tails due to resolution effects.

The event-level efficiency for staus to pass a variety of $\beta_{\mathrm{TOF}}$ and $m_{\mathrm{TOF}}$ requirements are shown in Figure~\ref{fig:hscp_delay_eff} of Appendix~\ref{app:hscp}. Efficiencies are shown for stable staus with respect to having at least one track in geometric acceptance. The selections on $\beta_{\mathrm{TOF}} < 0.96, 0.95, 0.90$ or $m_{\mathrm{TOF}} >15, 30, 60$~GeV were chosen to provide comparable background efficiencies of $10\%$, $5\%$, and $1\%$ per track, respectively. In both cases, the event-level efficiency increases with stau mass. Requirements on the measured $\beta_{\mathrm{TOF}}$ are less efficient than the measured mass.

\FloatBarrier
\section{Comparison of tracking parameters}
\label{sec:comparisons}

The effect of varying primary tracking parameters is compared across BSM models and signatures in the following sections. The minimum \pt threshold is relevant to some extent for all models, primarily affecting SUEPs and the Higgs portal model. Maximum reconstructed \dz affects only models with displaced tracks. The final factor considered is the detector layout, as studied through the HSCP model.

\subsection{Minimum transverse momentum}

\begin{figure}[htb]
     \centering
     \begin{subfigure}[b]{0.49\textwidth}
         \centering
        \includegraphics[width=\textwidth]{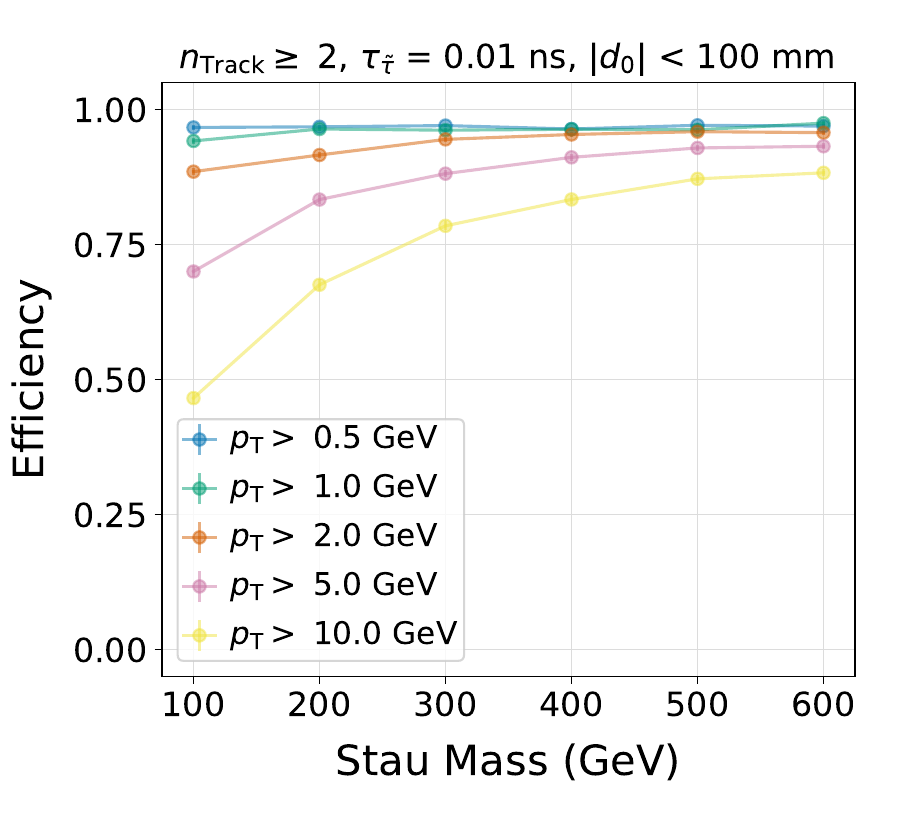}
         \caption{ }
         \label{subfig:stau_pt_comp}
     \end{subfigure}
     \hfill
     \begin{subfigure}[b]{0.49\textwidth}
         \centering
        \includegraphics[width=\textwidth]{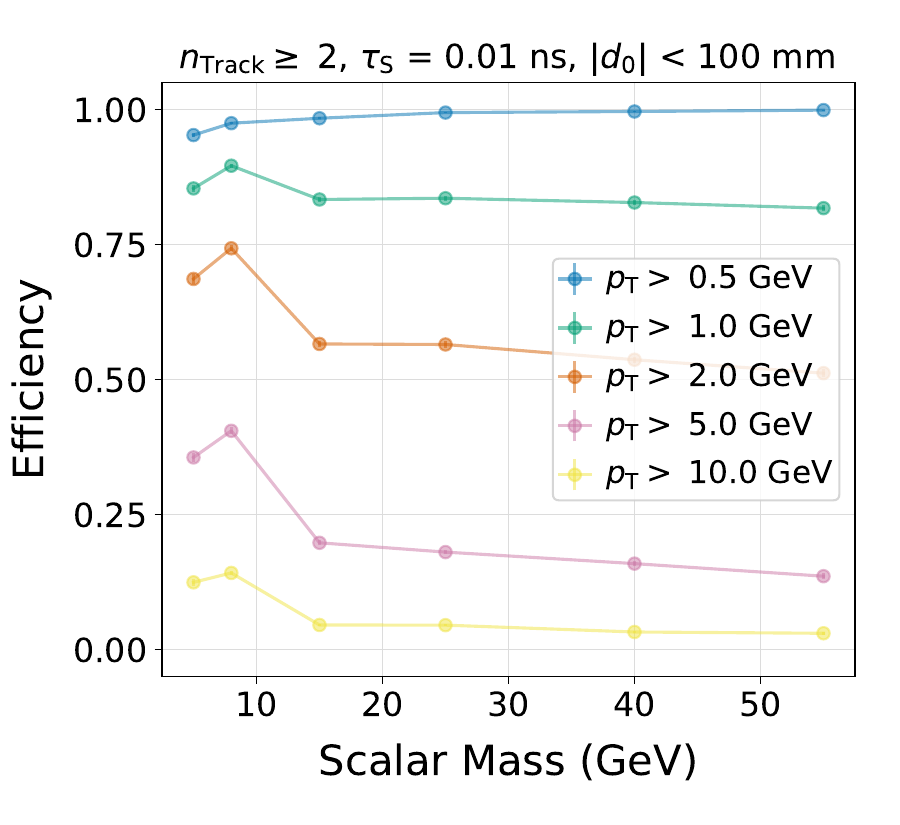}
         \caption{ }
         \label{subfig:higg_pt_comp}
     \end{subfigure}
     
     \begin{subfigure}[b]{0.49\textwidth}
         \centering
        \includegraphics[width=\textwidth]{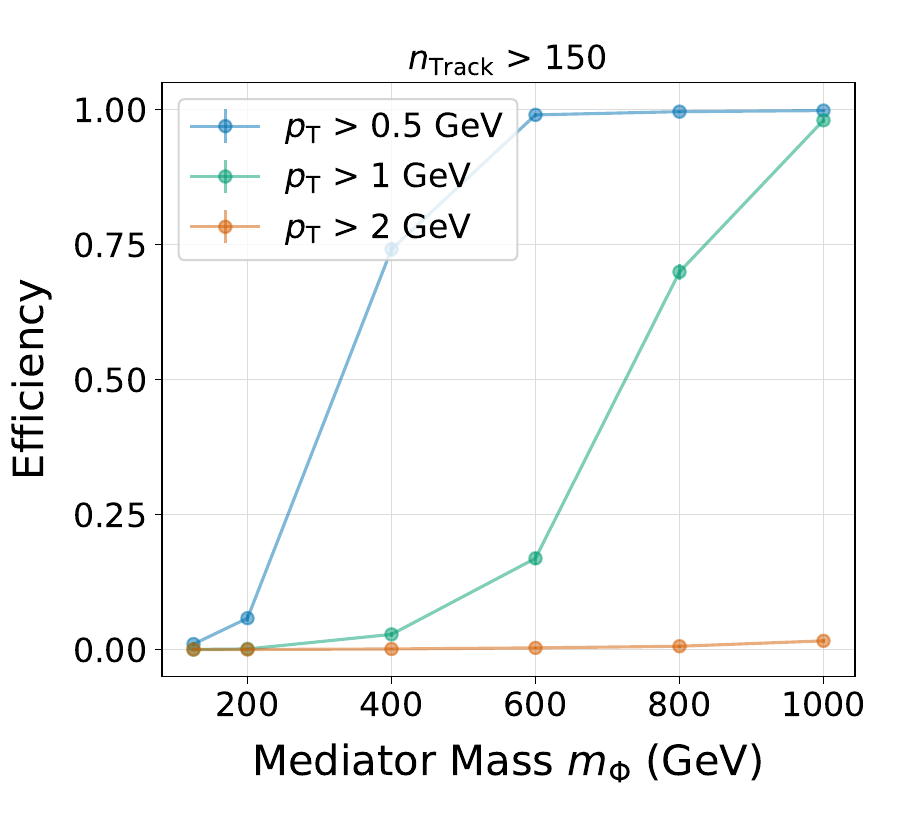} 
         \caption{ }
         \label{subfig:suep_pt_comp}
     \end{subfigure}     
    \caption{Trigger efficiency as a function of BSM particle mass for the stau (\subref{subfig:stau_pt_comp}), Higgs portal (\subref{subfig:higg_pt_comp}), and SUEPs (\subref{subfig:suep_pt_comp}) models for a range of minimum track \pt thresholds. The stau and Higgs portal samples both correspond to a lifetime of $\tau =0.01$~ns in order to target displaced decay products. A linearly decreasing efficiency is assumed up to a $\dz$-endpoint of $100$~mm, and both long-lived scenarios require $\nTrack \geq 2$. In the SUEP model all tracks are prompt.}
    \label{fig:pt_comparisons}
\end{figure}

Event-level efficiencies as a function of the minimum track \pt threshold and BSM parent mass are shown for several different models in Figure~\ref{fig:pt_comparisons}. The \pt threshold has a negligible effect on the direct detection of most stable charged particles, and is not considered here. For this comparison, the Higgs portal trigger selection is loosened to consider an event accepted if at least two, rather than five, tracks are successfully reconstructed. This adjustment makes the two selections identical and the results directly comparable.

Displaced leptons, like heavy stable charged particles, are relatively robust to the \pt threshold, though efficiencies would drop below $50\%$ for the lightest staus with a \pt$ > 10$ GeV requirement. The \pt threshold has a larger effect on the Higgs portal scenario, where the hadronic decay products are usually softer. Thresholds above 2 GeV would reduce the efficiency of even this two-track trigger to below $20\%$ for all Higgs portal scalar masses. The largest impact of the \pt threshold is on the SUEP signature, where all masses have negligible efficiency for $\pt \geq 2$~GeV.

In most systems the reconstruction for prompt and displaced tracks is likely to be controlled and parameterized separately. In these cases, it may be possible to set a lower \pt threshold for prompt track reconstruction than for displaced tracks. This would enable triggering on prompt SUEP scenarios, where a low momentum requirement is the most essential. However, the \pt threshold for displaced tracks would still need to be on the order of $2$ to $3$~GeV to achieve significant efficiency for the Higgs portal model.

It should be noted that while the minimum track reconstruction \pt threshold should be kept as low as possible within the balance of other factors, individual triggers tuned for particular BSM scenarios may benefit from using a much higher minimum \pt requirement to suppress background. For example, the efficiency for detection of HSCPs was found to be nearly independent of minimum track \pt for cut values up to $100$~GeV.

\subsection{Impact parameter dependence}
\label{subsec:dzero}

\begin{figure}[htb]
     \centering
     \begin{subfigure}[b]{0.49\textwidth}
         \centering
        \includegraphics[width=\textwidth]{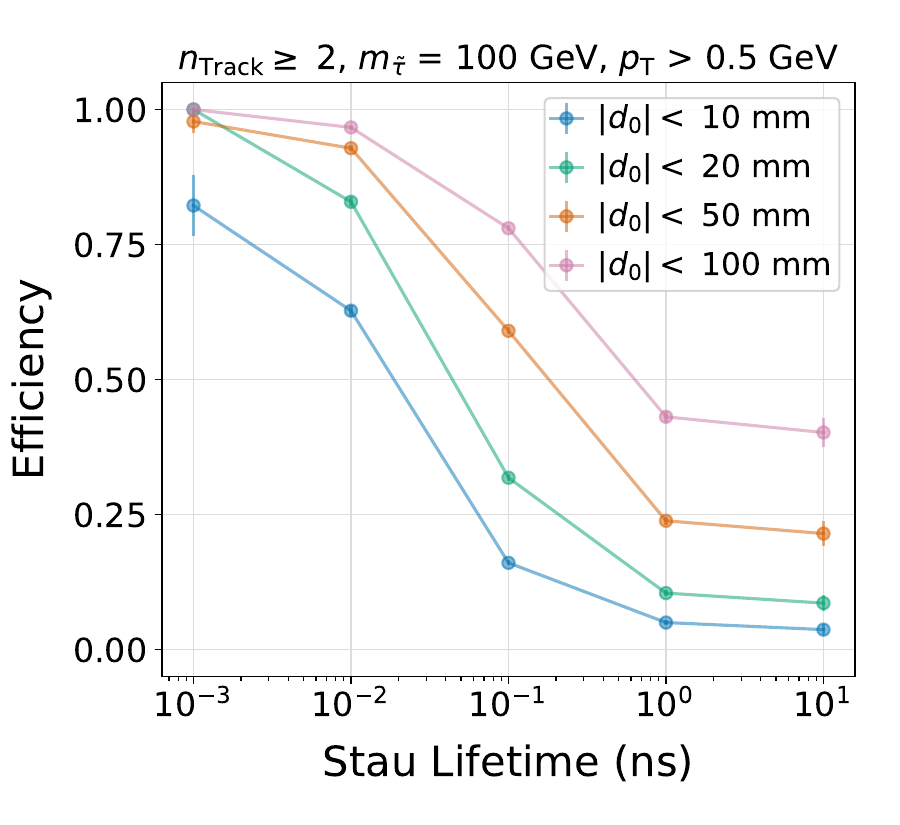}
         \caption{}
         \label{subfig:stau_d0_comp_1}
     \end{subfigure}
     \hfill
     \begin{subfigure}[b]{0.49\textwidth}
         \centering
        \includegraphics[width=\textwidth]{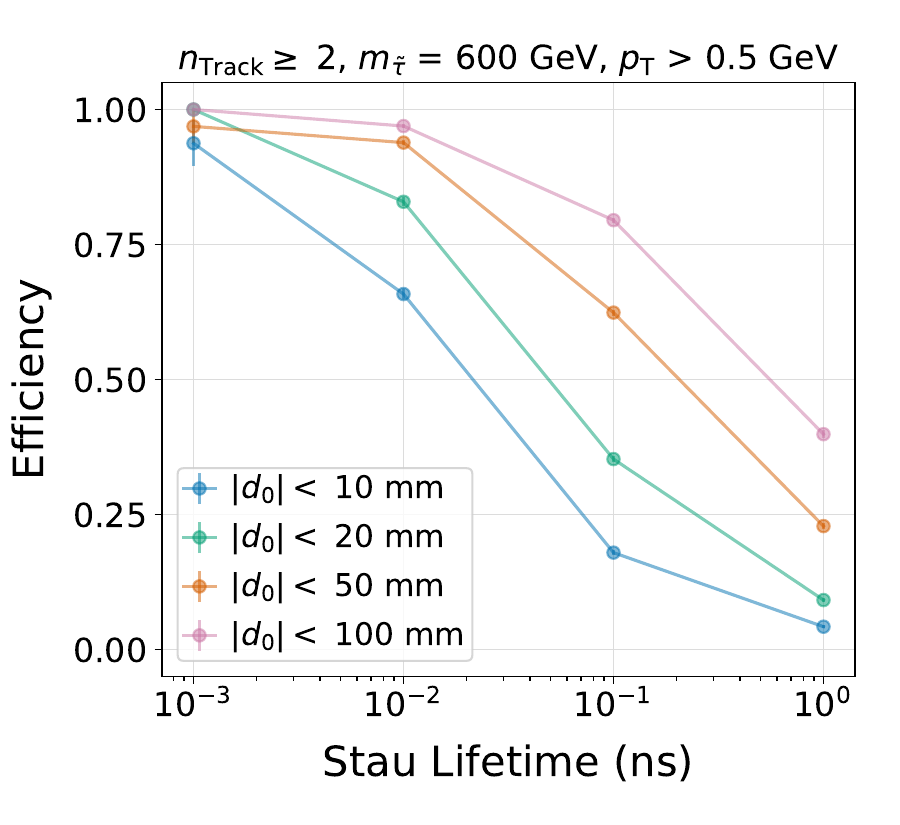}
         \caption{}
         \label{subfig:stau_d0_comp_2}
     \end{subfigure}
     
     \begin{subfigure}[b]{0.49\textwidth}
         \centering
        \includegraphics[width=\textwidth]{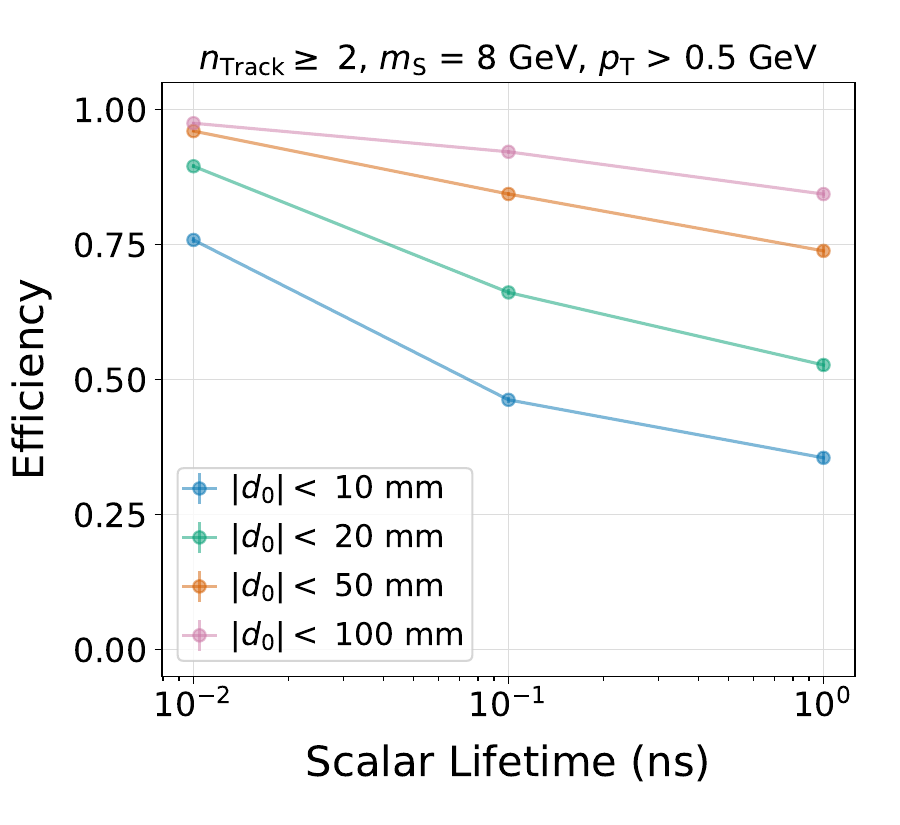}
         \caption{}
         \label{subfig:higg_d0_comp_1}
     \end{subfigure}
     \hfill
     \begin{subfigure}[b]{0.49\textwidth}
         \centering
        \includegraphics[width=\textwidth]{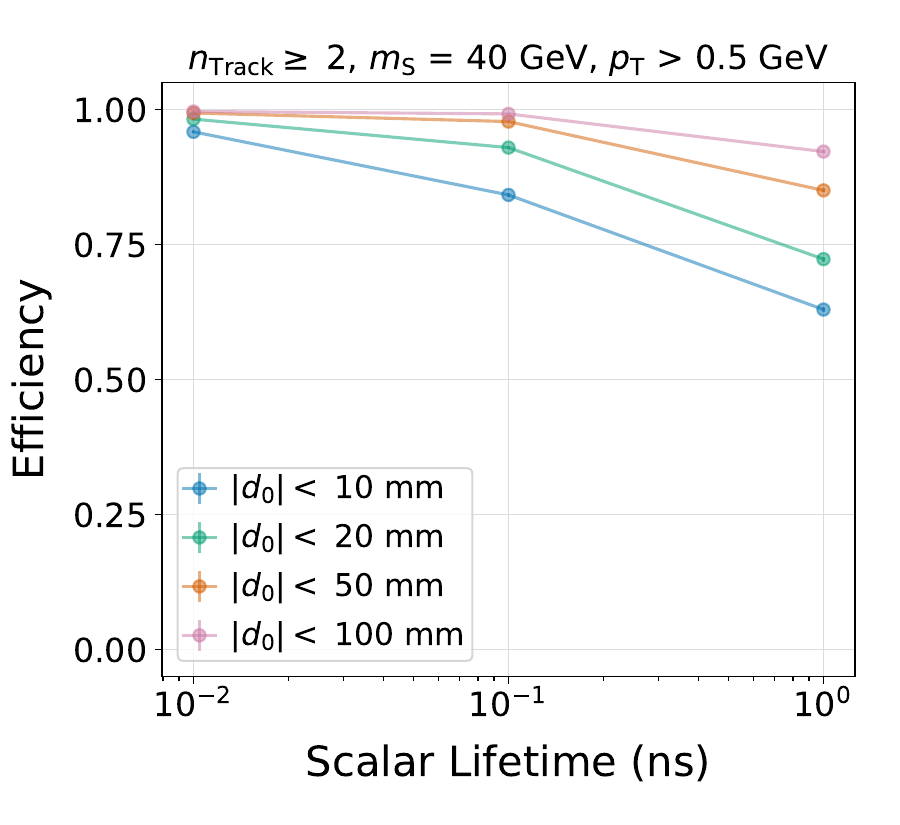}
         \caption{}
         \label{subfig:higg_d0_comp_2}
     \end{subfigure}      
    \caption{Trigger efficiency as a function of lifetime at a range of maximum track \dz values for staus with masses 100 GeV (\subref{subfig:higg_d0_comp_1}) and 600 GeV (\subref{subfig:stau_d0_comp_2}) and a Higgs portal with scalar masses 8 GeV (\subref{subfig:higg_pt_comp}) and 40 GeV (\subref{subfig:higg_d0_comp_2}). Only displaced tracks produced in the long-lived particle decay are considered. Effects of the \pt threshold are minimized by requiring $\pt>0.5$~ GeV, and both long-lived scenarios require $\nTrack \geq 2$.
    } 
    \label{fig:d0_comparisons}
\end{figure}

Trigger efficiency dependence on the maximum reconstructed \dz is shown as a function of long-lived particle lifetime for the stau and Higgs portal models in Figure~\ref{fig:d0_comparisons}. As is the case in the \pt comparison, the Higgs portal scenario requires $\nTrack \geq 2$, rather than five. The requirements for an event to pass are therefore identical between these two models. Two mass points for each scenario are chosen to illustrate the full range of displaced track kinematics. For the stau model, one low and one high mass point are chosen. For the Higgs portal, mass points are chosen such that scalar decays are either primarily to $b$ quarks or to $c$ quarks and taus.

\begin{figure}[htb]
     \centering
     \begin{subfigure}[b]{0.49\textwidth}
         \centering
        \includegraphics[width=\textwidth]{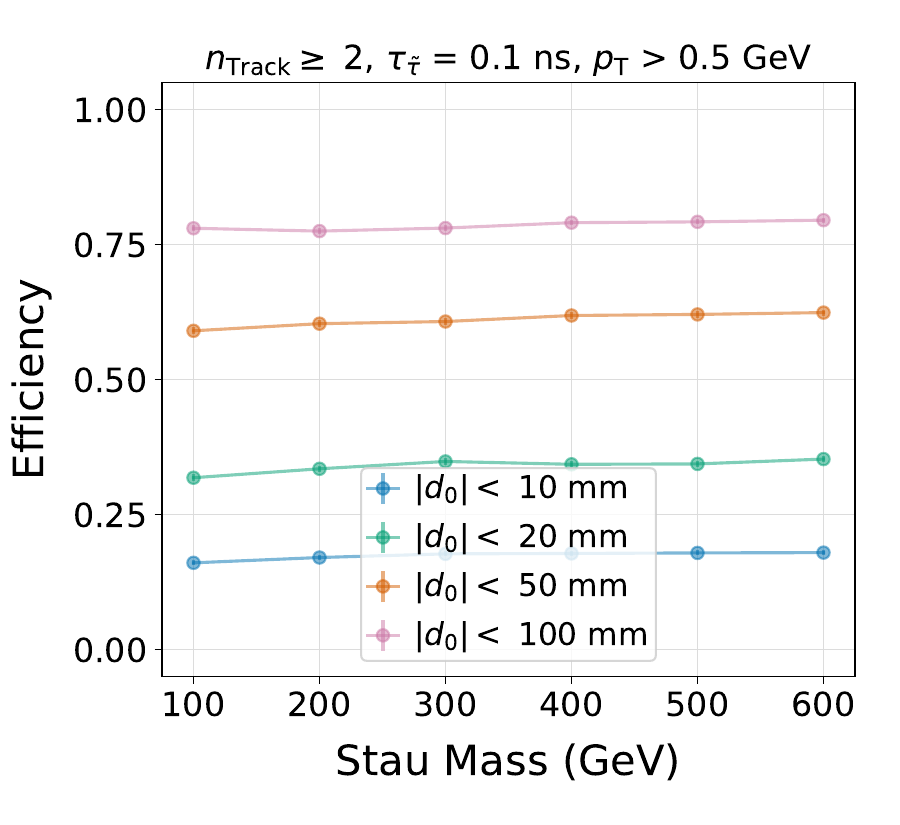}
         \caption{}
         \label{subfig:vary_d0_1}
     \end{subfigure}
     \hfill
     \begin{subfigure}[b]{0.49\textwidth}
         \centering
        \includegraphics[width=\textwidth]{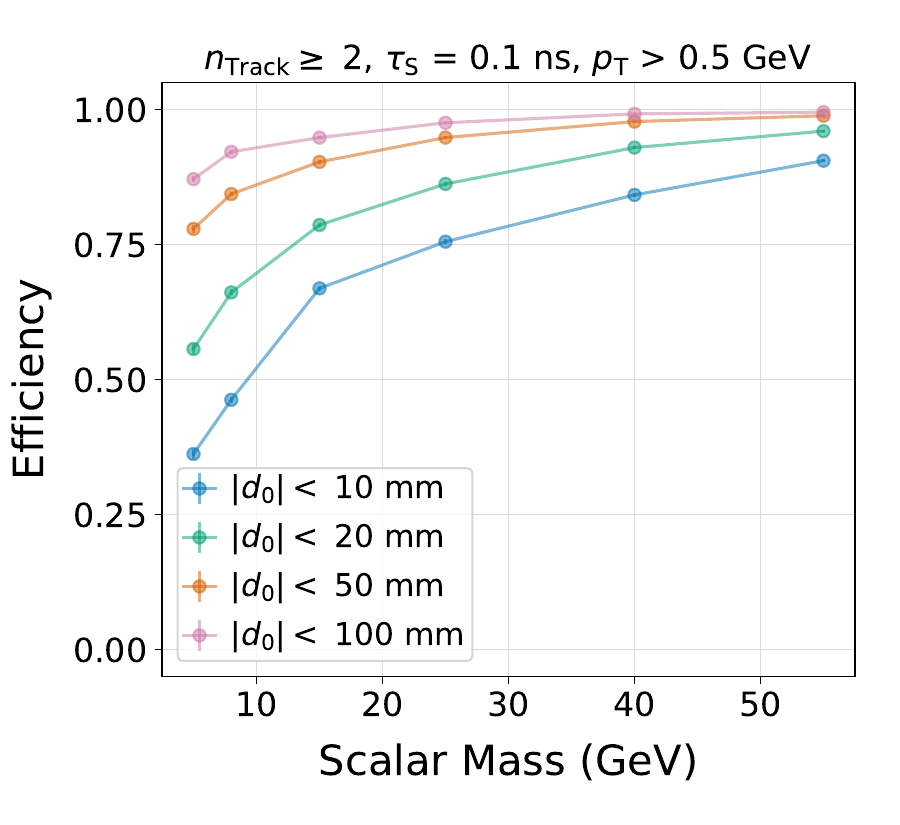}
         \caption{}
         \label{subfig:vary_d0_2}
     \end{subfigure}
\caption{Event-level efficiency for different $|\dz|$-endpoints as a function of long-lived particle mass. Both the stau (\subref{subfig:vary_d0_1}) and  Higgs portal  model (\subref{subfig:vary_d0_2}) have lifetimes of $\tau=0.1$~ns. Only displaced tracks produced in the long-lived particle decay are considered. Effects of the \pt threshold are minimized by requiring $\pt>0.5$~ GeV, and both long-lived scenarios require $\nTrack \geq 2$.
}
\label{fig:vary_d0}
\end{figure}

The trends in Figure~\ref{fig:d0_comparisons} demonstrate that the largest change in efficiency as a result of reduced $|\dz|$-endpoints occurs for intermediate lifetimes ($10^{-2} \lesssim \tau \lesssim 1$~ns) for both benchmark models. For proper lifetimes of $\tau=0.1$~ns, Figure~\ref{fig:vary_d0} shows the impact of varying the \dz-endpoint for a variety of stau and scalar masses. The difference in efficiency for different endpoints is consistent for the range of long-lived particle masses considered. 

Figure~\ref{fig:vary_d0} also shows that the impact of \dz-endpoints is significantly more important for the stau model than for the Higgs portal model. This effect is due to a difference in boost. The long-lived scalar's decay products are more collimated and often point back to the $pp$-collision. As a result, displaced tracks in the Higgs portal model tend to have smaller \dz for a given long-lived particle decay position, with respect to the stau model. It is therefore possible to achieve comparable efficiencies between tight and extended \dz-endpoints in the Higgs portal model. In contrast, the less boosted stau model suffers major efficiency losses as the \dz-endpoint is reduced from $10$~cm to $10$~mm.

\FloatBarrier
\subsection{Number of tracking detector layers}

The minimum number of tracking detector layers required per track was studied for the heavy metastable charged particle scenario by requiring the stau to decay beyond a varying minimum $\Lxy$. Reducing the minimum $\Lxy$ from $1.2$ to $0.6$~m was not found to have a significant effect on the event-level acceptance. For staus with a lifetime of $\tau_{\tilde{\tau}}=1$~ns, the difference between the most stringent and least stringent requirements on the effective number of tracker layers led to a factor four improvement in event-level acceptance. However, the acceptance remained low in all cases. For even longer lifetimes, most staus decay well beyond $\Lxy \sim 1$~m. Reducing the minimum number of layers required per track resulted in minimal acceptance improvements.

The effects of the number of tracking layers will still be significant for displaced signatures. The radius at which a long-lived particle decays will directly affect how many tracking layers are available to reconstruct the displaced track. For example, if the particle decays after the first tracking layer, only $n-1$ hits will be associated to the reconstructed track. The long-lived particle's decay distance also correlates to $\dz$, with the mean decay distance longer than the mean $\dz$. If a long-lived particle decays beyond multiple tracker layers, there may not be enough hits to reconstruct displaced tracks. The efficiency for reconstructing tracks beyond the equivalent \dz would drop to zero.

As seen in the Section~\ref{subsec:dzero}, access to high \absdz tracks significantly increases the efficiency for both stau and Higgs portal models. For staus, reconstructing high \absdz tracks enables the displaced track trigger to outperform the stable charged particle trigger at a lifetime of $1$~ns. For the stable charged particle approach, staus with masses between $100$~GeV and $500$~GeV at $1$~ns lifetimes all show an acceptance times efficiency below $15\%$ regardless of the number of layers required per track. For the displaced track approach, the acceptance times efficiency for a stau with a $1$~ns lifetime increases from $\sim 0\%$ to $\sim 20\%$ if the \absdz-endpoint is extended from $10$~mm to $100$~mm, assuming a linearly decreasing efficiency (see Appendix~\ref{app:displaced_lep}). Extending the efficiency to a larger \absdz-endpoint relies on having a sufficient number of tracking layers at larger radii.

\FloatBarrier
\section{Conclusions and Recommendations}

Current LHC triggers are primarily designed to target prompt decays of heavy particles to high momentum Standard Model particles. However, many compelling BSM scenarios lead to decays which are significantly displaced from the interaction point or include low momentum decay products. For these unconventional scenarios, the anomalous charged particle trajectories are the most distinctive feature of the event. This study explored the impact of track reconstruction parameters on trigger efficiency for a collection of representative models.  

In general, for these unconventional signatures, perfect track reconstruction efficiency is rarely achievable. However, even introducing a partial track reconstruction efficiency in a \pt or \dz range can make a substantial difference in sensitivity, and high efficiencies across these parameter spaces are not required. Triggering on $20\%$ of events is infinitely better than triggering on none.

HSCP sensitivity is relatively insensitive to choice of tracker parameters, due to their distinctive, high-\pt tracks. If timing information is available, translating this measurement into a time of flight estimate is the most discriminating strategy. For HSCPs, the most impactful variation studied was $\eta$ coverage, with typical efficiencies from a barrel-only scenario at $\sim 30\%$, but those from an extended endcap scenario as high as $\sim 75\%$.

To tackle highly diffuse models such as SUEPs, a track trigger with a low \pt-threshold is required. For the models studied here, the highest possible useful threshold is $\pt > 1$~GeV, though variations in model temperature would affect this cut-off. For experiments where a low \pt-threshold is not feasible at L1, alternatives to track counting should be considered, and could be complemented with lower \pt-thresholds at HLT. For the SUEP scenario considered here, the overall efficiency is still likely to be low, because high-energy initial state radiation is required for any alternate trigger strategy. If the \pt-threshold could be kept at $1$~GeV or pushed lower for prompt tracks, it would be possible to drastically improve sensitivity to SUEPs.

For the other models considered here, a tracking threshold of $\pt >2$~GeV is sufficient to provide useful sensitivity. In the case of the Higgs portal model, substantial improvement could be gained from reduced \pt thresholds. To address models with light mediators, lower mass long-lived particles, and hadronic decays, the most useful design strategy is a track trigger that pushes the \pt threshold as low as possible with modest \dz coverage. To gain sensitivity to the decay products of more massive particles, such as long-lived staus, a large \dz range is more essential while a low \pt threshold is not. 

To address all of these cases simultaneously while keeping resource usage to a minimum, the best design scenario is one that allows for increasingly large \dz coverage as particle \pt increases. In track trigger designs using pattern banks, this strategy is relatively easy to employ. An approximately constant number of patterns is required to cover a given area in the 2-dimensional space defined by maximum \dz and $1/\pt$ values. An optimal pattern bank could be defined sampling from a triangle rather than a rectangle in that 2D space. A more generic strategy is to create a displaced tracking algorithm that is applied only to tracks passing a minimum \pt threshold that is higher than the nominal prompt one. For example, a tracker with prompt tracking for tracks with $\pt >$~1 GeV and displaced tracking for $\pt > 2$~GeV would provide new sensitivity to all models considered here. 

Track triggers at the High Luminosity LHC will provide a unique opportunity to directly trigger on these challenging exotic scenarios for the first time. Acting on the recommendations provided here will enable track trigger designs for the HL-LHC to strike a balance that can best serve a full range of unconventional signatures. The same principles will apply to tracker and trigger designs for future detectors.

\acknowledgments

Tova Holmes and Jesse Farr's research is supported by U.S. Department of Energy, Office of Science, Office of Basic Energy Sciences Energy Frontier Research Centers program under Award Number DE-SC0020267.

Karri Folan DiPetrillo and Chris Guo's work utilizes resources of the Fermi National Accelerator Laboratory (Fermilab), a U.S. Department of Energy, Office of Science, HEP User Facility. Fermilab is managed by Fermi Research Alliance, LLC (FRA), acting under Contract No. DE-AC02-07CH11359. 

Jess Nelson's research is supported by the Duke Research Experiences for Undergraduates program in particle physics, under National Science Foundation Award Number 1757783. 

Katherine Pachal's research is supported by TRIUMF, which receives federal funding via a contribution agreement with the National Research Council (NRC) of Canada.









\FloatBarrier
\printbibliography

\appendix
\section{Additional Distributions}

\subsection{Higgs portal}\label{app:higgs_portal}

\begin{figure}[htbp]
\centering 
\includegraphics[width=0.48\textwidth]{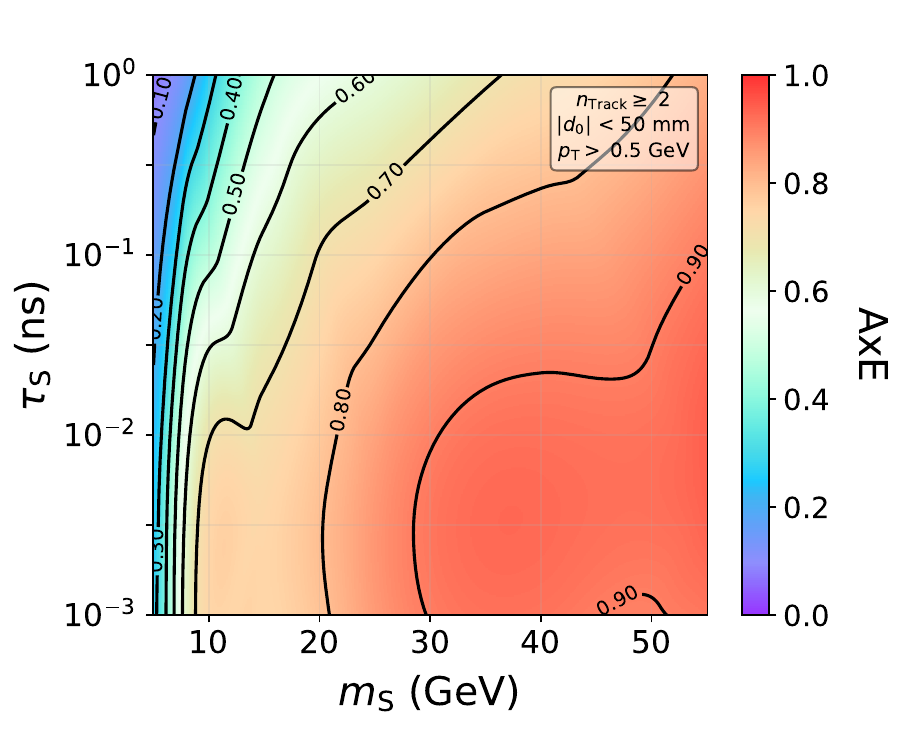} \hfill
\includegraphics[width=0.48\textwidth]{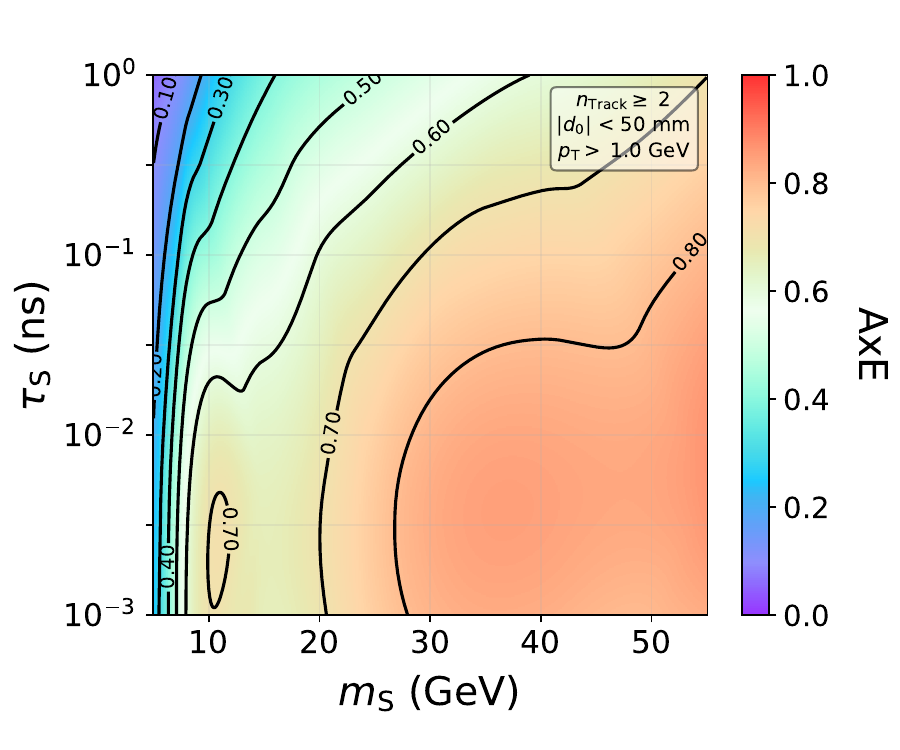} \\
\includegraphics[width=0.48\textwidth]{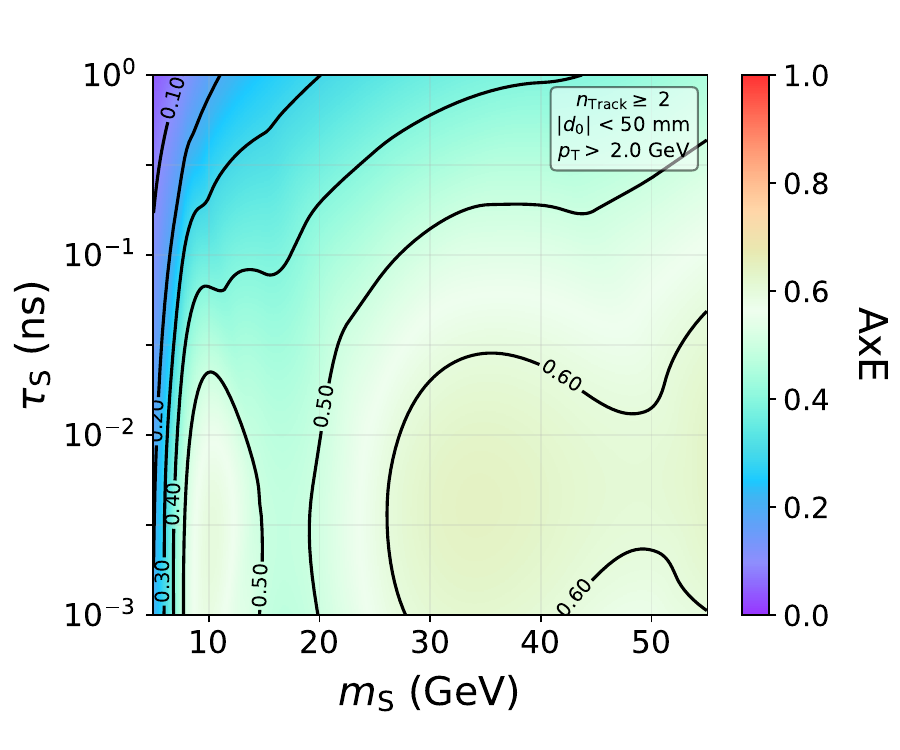} \hfill
\includegraphics[width=0.48\textwidth]{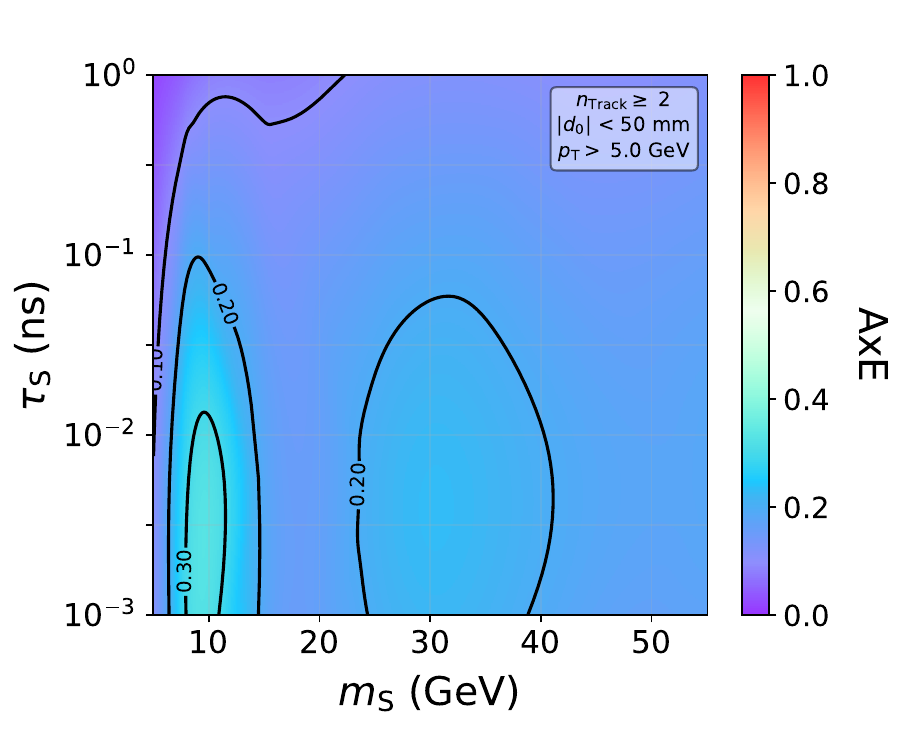}
\caption{\label{fig:higgs_acctimeseff} Event-level acceptance times efficiency for Higgs portal model with a constant \dz-endpoint and \pt-thresholds ranging (clockwise from top left to bottom left) from 0.5 Gev, 1.0 GeV, 2.0 Gev, and 5.0 GeV. In all cases, the two-track selection and a linearly decreasing efficiency is used.}
\end{figure}

\FloatBarrier
\subsection{Displaced Leptons}\label{app:displaced_lep}

\begin{figure}[htbp]
\centering 
\includegraphics[width=0.48\textwidth]{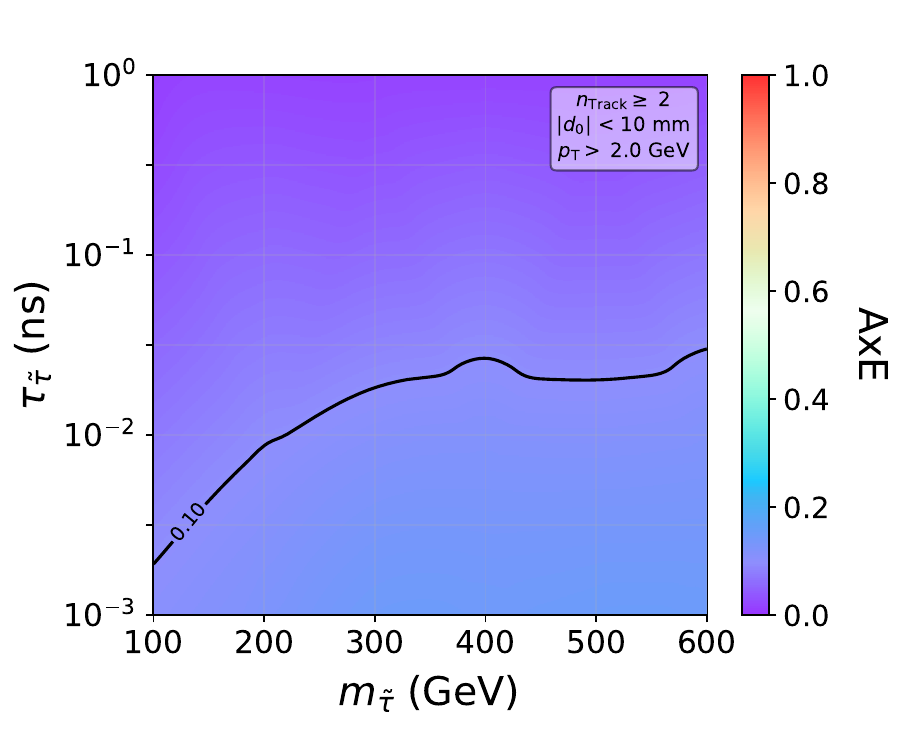}\hfill
\includegraphics[width=0.48\textwidth]{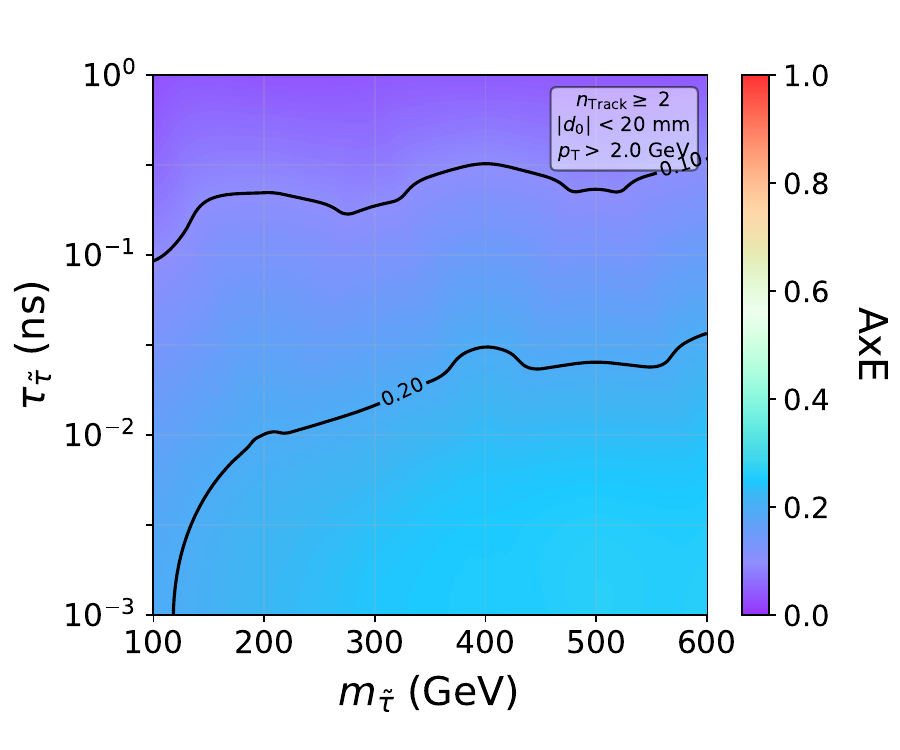} \\
\includegraphics[width=0.48\textwidth]{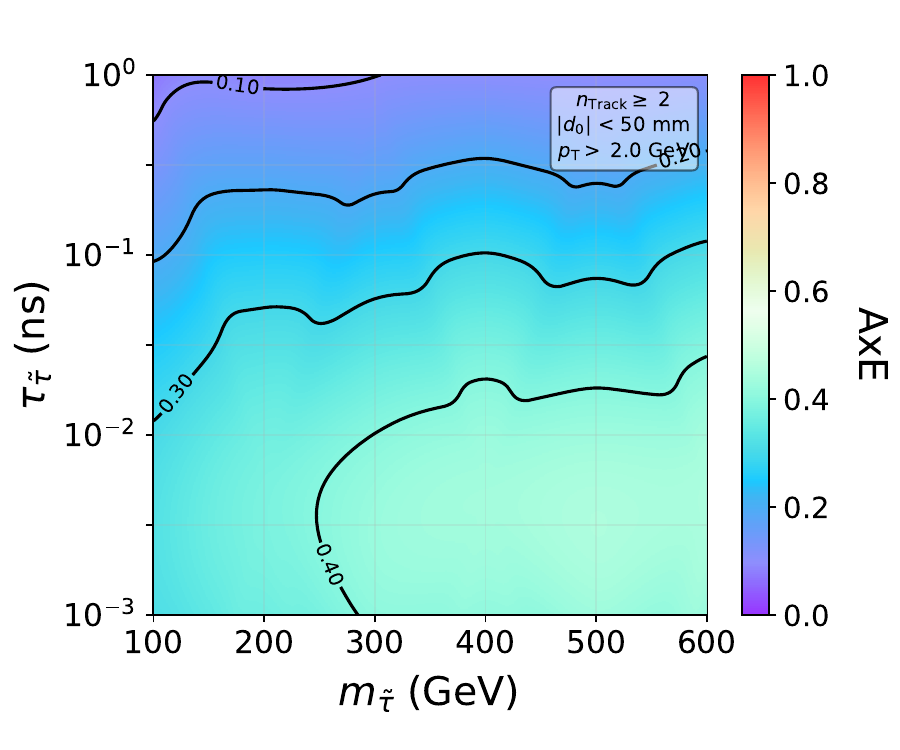}\hfill
\includegraphics[width=0.48\textwidth]{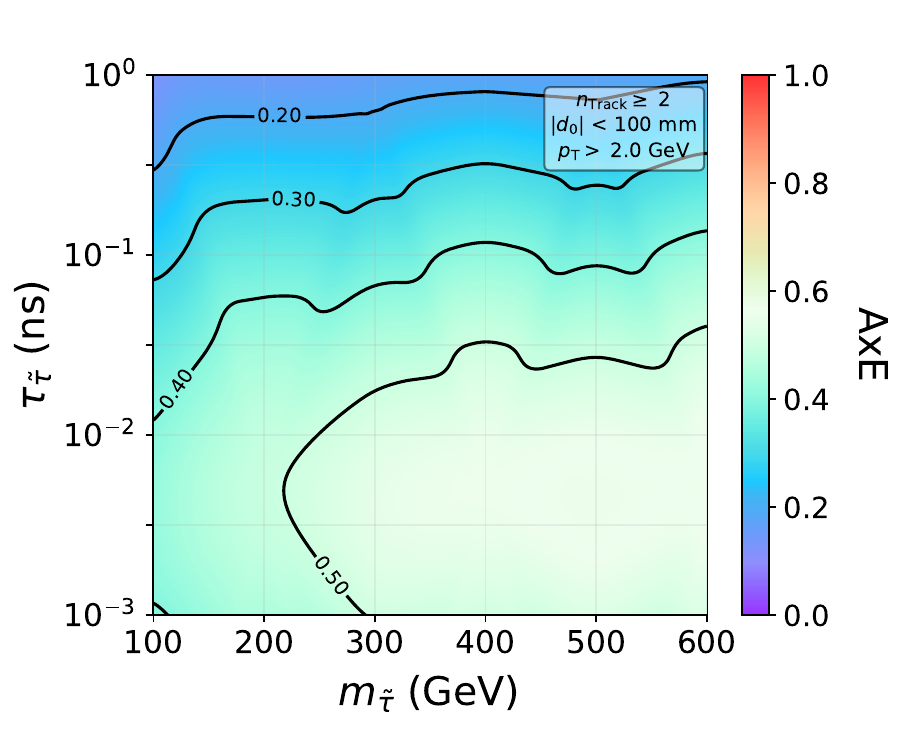}
\caption{\label{fig:stau_acctimeseff} Event-level acceptance times efficiency for a displaced track trigger targeting the stau scenario. The constant \pt-threshold is kept constant and \dz-endpoing is varied from (clockwise from top left to bottom left) from 10 mm, 20 mm, 50 mm, and 100 mm. In all cases, the two-track selection and a linearly decreasing efficiency is used. 
}
\end{figure}


\FloatBarrier
\subsection{HSCP}\label{app:hscp}

\begin{figure}[htbp]
\centering 
\includegraphics[width=.6\textwidth]{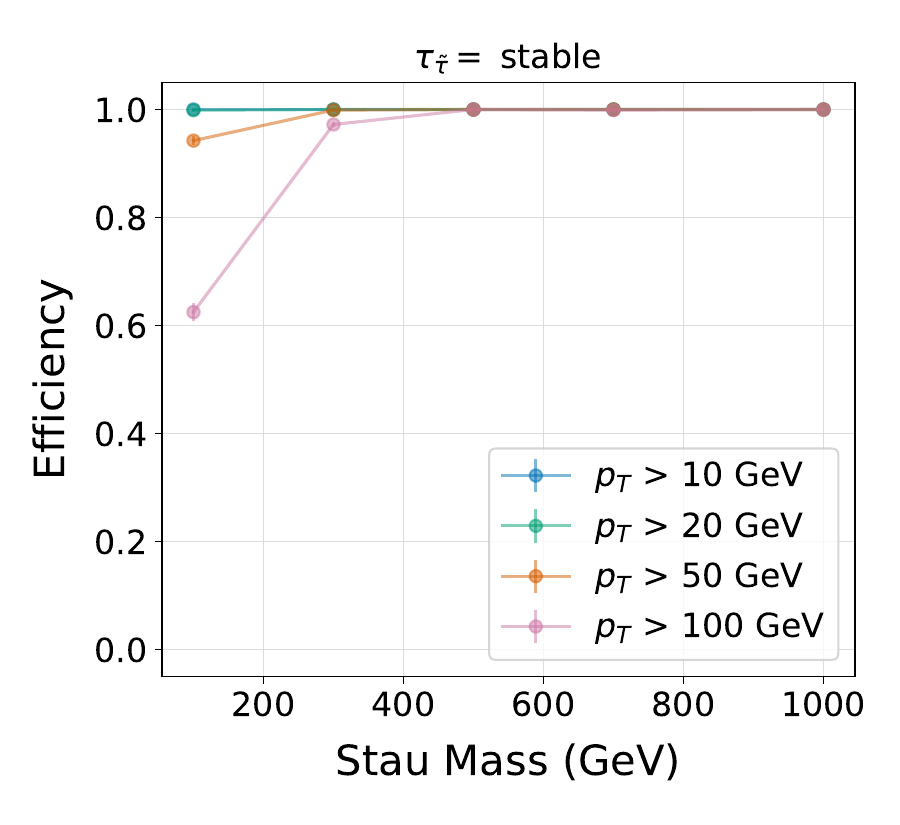}
\caption{\label{fig:hscp_isolatedtrack_pt} Stau efficiency for a variety of minimum tracker \pt values. Efficiency is shown as a function of stau mass for a stable lifetime and minimum $\Lxy=1200$~mm. }
\end{figure}

\begin{figure}[htb]
     \centering
     \begin{subfigure}[b]{0.49\textwidth}
         \centering
         \includegraphics[width=\textwidth]{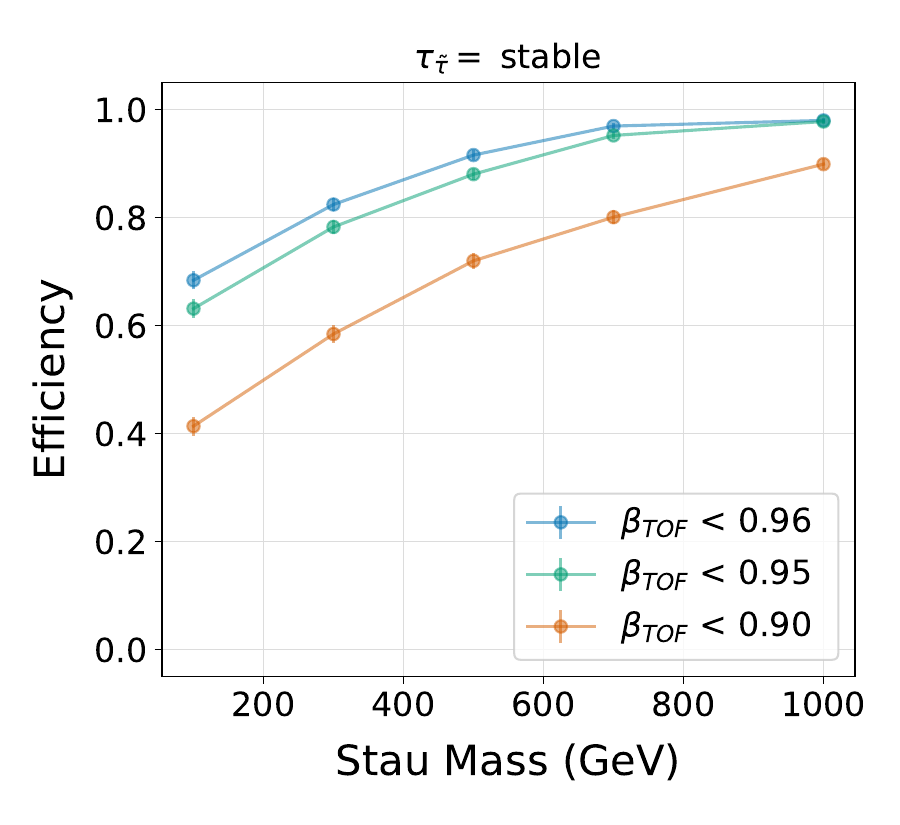}
         \caption{ }
         \label{subfig:hscp_eff_beta}
     \end{subfigure}
     \hfill
     \begin{subfigure}[b]{0.49\textwidth}
         \centering
         \includegraphics[width=\textwidth]{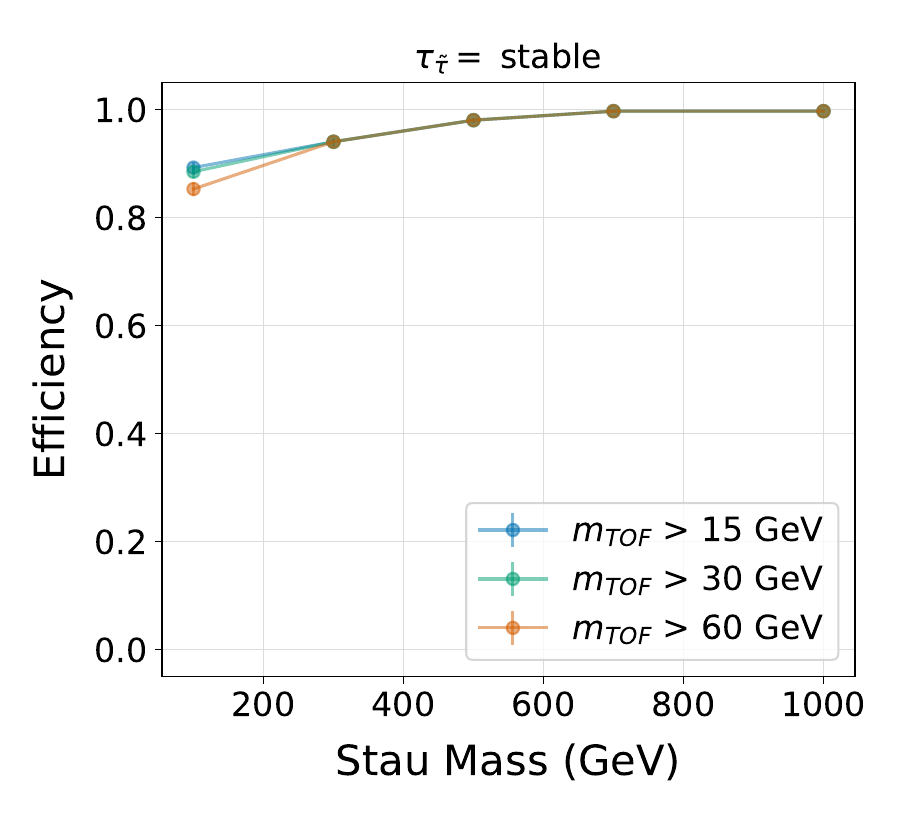}
         \caption{ }
         \label{subfig:hscp_eff_mass}
     \end{subfigure}
    \caption{Efficiency with respect to various requirements on measured track $\beta$ (\subref{subfig:hscp_eff_beta}) and measured mass (\subref{subfig:hscp_eff_mass}).
    }
    \label{fig:hscp_delay_eff}
\end{figure}

\end{document}